\documentclass[aps,prd,preprint,showpacs,superscriptaddress,noshowpacs]{revtex4}
\usepackage{gensymb}
\usepackage{graphicx}  
\usepackage{dcolumn}   
\usepackage{bm}        
\usepackage{amssymb}   
\usepackage{xcolor}
\usepackage{graphicx}
\usepackage{pdfpages}
\usepackage{multirow}
\usepackage{amsmath}

\usepackage[caption=false,subrefformat=parens,labelformat=parens]{subfig}

\usepackage{tikz}
\usetikzlibrary{positioning,arrows}
\usetikzlibrary{decorations.pathmorphing}
\usetikzlibrary{decorations.markings}

\usepackage{hyperref}
\hypersetup{
    colorlinks,
    linkcolor={red!70!black},
    citecolor={blue!90!black},
    urlcolor={blue!90!black}
}

\usepackage{xspace}

\newcommand*{\ie}{\textit{i.e.,}\@\xspace}

\makeatletter
\newcommand*{\etc}{%
    \@ifnextchar{.}%
        {etc}%
        {\textit{etc.}\@\xspace}%
}
\makeatother

\begin{document}

\title{Study of neutrino oscillation parameters \\at the $\textsc{INO-ICAL}$ detector using event-by-event reconstruction}

\author{Karaparambil Rajan Rebin}
\email{rebin@physics.iitm.ac.in}
\affiliation{Indian Institute of Technology Madras, Chennai 600 031, India}
\author{Jim Libby}
\email{libby@iitm.ac.in}
\affiliation{Indian Institute of Technology Madras, Chennai 600 031, India}
\author{D. Indumathi}
\email{indu@imsc.res.in}
\affiliation{The Institute of Mathematical Sciences, Chennai 600 113, India}
\author{Lakshmi S. Mohan}
\email{lakshmilsm9@gmail.com}
\affiliation{Indian Institute of Technology Madras, Chennai 600 031, India}

\date{\today}
\begin{abstract}
We present the reach of the proposed INO-ICAL in measuring the atmospheric-neutrino-oscillation parameters $\theta_{23}$ and $\Delta m^2_{32}$ using full event-by-event reconstruction for the first time. We also study the fluctuations in the data and their effect on the precision measurements and mass-hierarchy analysis for a five-year exposure of the 50 kton ICAL detector. We find a mean resolution of $\Delta\chi^2 \approx 2.9$, which rules out the wrong mass hierarchy of the neutrinos with a significance of approximately $1.7\sigma$. These results are similar to those to presented earlier studies that approximated the performance of the ICAL detector.
\end{abstract}


\maketitle

\section{Introduction\label{1}}
\paragraph*{}
In the Standard Model (SM), neutrinos are massless fermions which interact only via the weak interaction through the exchange of $W^\pm$ or $Z^0$ bosons. A series of experiments dedicated to neutrinos \cite{SOL_SNO, ATM_SK, KAM_08, K2K_06, T2K_11, CHOOZ_12, MINOS_12, RENO_12, DAYA_12} have proved the existence of neutrino flavour oscillations, which implies that neutrinos are massive. The neutrino flavour states produced along with charged leptons are linear superpositions of the mass eigenstates. Due to the difference in phase between the wave packets of each of the mass eigenstates, neutrino oscillations occur \cite{pon, maki}.

\paragraph*{}
In the case of three neutrino flavours, the mixing is described by a $3\times3$ unitary matrix called the PMNS (Pontecorvo-Maki-Nakagawa-Sakata) matrix \cite{pon, maki}, where the oscillations are governed by the following parameters: two mass-square difference terms $(\Delta m^{2}_{21}, \Delta m^{2}_{32};~\Delta m^2_{ij}\equiv m_i^2-m_j^2;~i,j=1,2,3;~i\neq j)$, mixing angles ($\theta_{12}, \theta_{13}, \theta_{23}$) and one $(\delta)$ or three $(\delta, \alpha_{21}, \alpha_{31})$ $CP$ violating phases, depending on whether neutrinos are Dirac or Majorana \cite{Majorana, Majorana_cp} particles, respectively. The mixing angle $\theta_{13}$, which determines the magnitude of $CP$-violation effects in neutrino oscillations, is found to be  non-zero from reactor \cite{CHOOZ_12, RENO_12, DAYA_12} and accelerator \cite{T2K_11p, MINOS_13} neutrino oscillation experiments probing  $\bar\nu_e$ disappearance and $\nu_e$ appearance, respectively. Solar neutrino oscillation parameters $\theta_{12}$ and $\Delta m^{2}_{21}$ have been measured combining data from solar neutrinos $(\nu_e)$  and KamLAND reactor neutrinos $(\bar\nu_e)$. The latest oscillation analysis of solar and KamLAND data \cite{KAM_new} gives $\Delta m^{2}_{21}=(7.37^{+0.17}_{-0.16})\times 10^{-5}~\mathrm{eV^2}$ and $\sin^2\theta_{12}=0.297^{+0.017}_{-0.016}$. Existing data from SK (Super-Kamiokande) \cite{sk_17}, T2K \cite{T2K_17}, MINOS \cite{MINOS_14} and NO$\nu$A \cite{NOVA_17} experiments give constraints on atmospheric neutrino oscillation parameters $|\Delta m^{2}_{32}|$ and $\theta_{23}$. However, the sign of $\Delta m^{2}_{32}$ and whether $\theta_{23}$ results in maximal mixing $\theta_{23}=45^{\circ}$ or is in the upper (lower) quadrant $\theta_{23}>45^{\circ}$ $(\theta_{23}<45^{\circ})$, is yet to be determined. The present information on these parameters can be found in Table~\ref{tab1}.

\begin{table}[!htb]
\caption{\label{tab1}The current best-fit values of neutrino oscillation parameters and their $3\sigma$ allowed ranges assuming normal (NH) and inverted (IH) neutrino mass hierarchies. The values are taken from Ref. \cite{PDG_16}. For CP phase, $\delta_{cp}$, at $3\sigma$ no physical values are disfavored, hence the $2\sigma$ range is given. Here $\Delta m^2 \equiv m_3^2-(m_2^2+m_1^2)/2$.}
\begin{ruledtabular}
\begin{tabular}{llll}
\multicolumn{2}{l}{{\centering}Parameter}  & best-fit value& {\centering\arraybackslash} $3\sigma$ range\\
\hline
 \multirow{2}{*}{$\sin^2\theta_{23}$}& (NH) & $0.437$ & $0.379~-~0.616$\\
 &(IH) &$0.569$ & $0.383~-~0.637$ \\
 \multirow{2}{*}{$\sin^2\theta_{13}$}& (NH) & $0.0214$ & $0.0185~-~0.0246$\\
 &(IH) &$0.0218$ & $0.0186~-~0.0248$ \\
 $\sin^2\theta_{12}$& & $0.297$ & $0.250~-~0.354$\\
 $\Delta m^2_{21}$ $[10^{-5}~\mathrm{eV^2}]$& & $7.37$ & $6.93~-~7.97$\\
 \multirow{2}{*}{$\Delta m^2$~~$[10^{-3}~\mathrm{eV^2}]$}& (NH) & $2.50$ & $2.37~-~2.63$\\
 &(IH) &$2.46$ & $2.33~-~2.60$ \\
 
  \multirow{2}{*}{$\delta_{cp}$~~~~~$[\mathrm{rad}]$}& (NH) & $1.35$ & $0.92~-~1.99$ $(2\sigma)$\\
 &(IH) &$1.32$ & $0.83~-~1.99$ $(2\sigma)$ \\
\end{tabular}
\end{ruledtabular}
\end{table}

\paragraph*{}
The relatively large value of $\theta_{13}\cong8.6\degree$ has intensified the search for $CP$ violation effects in neutrino oscillations, and also the determination of the sign of $\Delta m^{2}_{32}$ via matter effects \cite{wolf, MSW}. Matter plays an important role in enhancing the effect of $\sin\theta_{13}$ via resonance, which is sensitive to the sign of $\Delta m^{2}_{32}$ and is different for neutrinos and antineutrinos \cite{Indu_prem, Blennow_matter}. Determination of the sign of $\Delta m^{2}_{32}$ would help us understand the correct mass hierarchy (MH) of the neutrinos, \ie whether the MH is $(m_1<m_2<m_3)$ normal (NH) or $(m_3<m_1<m_2)$ inverted (IH) hierarchy. A series of experiments with complementary approaches have been proposed using accelerator, reactor and atmospheric neutrinos to determine the MH \cite{Wor_grp_neutrinos, deter_MH}. The intermediate and long baseline, off-axis accelerator neutrino experiments T2K \cite{T2K_11} and NO$\nu$A \cite{NOVA_02, NOVA_13}, search for the appearance of $\nu_e$ in an intense beam of $\nu_\mu$, wherein the appearance probability depends on the MH of the neutrino states. Liquid scintillator detectors proposed at RENO-50 \cite{RENO_12} and JUNO \cite{JUNO} could unravel the MH using reactor neutrinos. Atmospheric neutrino experiments using water or ice Cherenkov  detectors, such as Hyper-K \cite{HK_11, HK_pro_13}, MEMPHYS \cite{MEMPHYS}, ORCA  and PINGU \cite{PINGU_13, PINGU_pro_13}, make use of different cross-sections and different $\nu$ and $\bar\nu$ fluxes to study the MH.

\paragraph*{}
The proposed magnetized Iron Calorimeter (ICAL), to be built at the India-based Neutrino Observatory (INO) \cite{INO_phy}, will study interactions involving primarily atmospheric muon neutrinos and anti-neutrinos. It will consist of three identical modules, each with dimension $16~\mathrm{m} \times 16~\mathrm{m} \times 14.5 ~\mathrm{m}$ placed in a line and separated by a small gap of $20~\mathrm{cm}$.  Each module will consist of $151$ layers of $5.6$ cm thick iron plates interleaved with 4~cm air gap containing the active detector elements, glass Resistive Plate Chambers (RPCs).  This huge size of $48~\mathrm{m} \times 16~\mathrm{m} \times 14.5 ~\mathrm{m}$ magnetized detector, with a mass of $50$ kton, composed provides the target nuclei to achieve a statistically significant number of neutrino interactions within a reasonable time frame. One of the main goals of INO is to study the MH via earth matter effects, and to determine the octant of $\theta_{23}$. ICAL is designed to have very good muon detection efficiency of greater than $85\%$ for muons greater than $2\mathrm{GeV}$ (with incident angle $\cos\theta\geq 0.4$), combined with excellent angular resolution. The most important property of the ICAL will be its ability to discriminate charge using the magnetic field. Thus, the ICAL can distinguish between $\nu_{\mu}$ and $\bar\nu_{\mu}$ events by observing the charge of the final state muons. Hence the ICAL could study the MH by observing earth-matter effects independently on $\nu$ and $\bar\nu$.

This paper shows the precision reach of ICAL in the $\sin^2\theta_{23}-|\Delta m^{2}_{32}|$ plane for a five year run of ICAL. Event-by-event reconstruction and fluctuations arising from low event statistics, using an analysis technique that will be suitable to be employed on the actual data. Previous studies \cite{INO_phy} used parameterizations of the efficiency and resolution that do not reflect the tails of these distributions, as well as using very large sample sizes to negate the effect of low statistics. We also apply a few event selection criteria, as presented in Ref. \cite{INOcutp}, but within the framework of low event statistics and present its effect on the outcome of this analysis. This paper also compares the results from simulated unfluctuated data, by simulating data sets corresponding to five-years of data. 
\paragraph*{}
This paper is organized as follows. In Sec. \ref{sec2}, we outline the methodology describing the event detection in the INO detector and the software framework used to simulate and reconstruct the events in the detector. In Sec. \ref{sec3}, we describe the event generation and discuss the fluctuations in the data. We describe how the events are reconstructed as well as the event selection criteria applied to obtain a sample of events. We also describe the oscillation analysis including the Earth-matter effects and discuss how data collected by the ICAL are sensitive to the MH. We also describe the $\chi^2$ analysis and the binning scheme used, and also discuss the types of systematic uncertainties used in this analysis.
\paragraph*{}
In Sec. \ref{sec4}, we present the results of our simulated analysis, showing the reach of the ICAL for atmospheric oscillation parameters $\theta_{23}$ and $\Delta m^{2}_{32}$. Event selection reduces the event statistics, hence we also present the results with and without event selection to see its effect. We also discuss the effect of fluctuations on the precision measurements and show the different possible outcomes resulting from the low event statistics. We also discuss the results on the MH of  the neutrinos and the effects of fluctuations in determining it. Finally, in Sec. \ref{sec5} we present the summary of our results and conclusions.
\section{Methodology}\label{sec2}
\paragraph*{}
The NUANCE \cite{NUANCE} neutrino event generator, along with the Honda neutrino flux \cite{HONDA} at the Kamioka site, is used to generate neutrino interactions within the ICAL detector. The proposed ICAL geometry, containing mainly iron and glass components of the detector, is given as input to NUANCE. It generates secondary particles from interactions with these materials, and calculates event rates integrated over the weighted flux and cross sections of all charged-current (CC) and neutral-current (NC) interactions, at each neutrino energy and angle. The output from NUANCE contains vertex and timing information, as well as energy and momentum of all initial and final state particles in each event.

In the ICAL, atmospheric neutrinos will interact with an iron nucleus, undergoing NC and CC interactions. The main CC interactions taking place in the detector are quasi-elastic (QE) and resonance (RS) at low energies and deep-inelastic scattering (DIS) at higher energies. All neutrinos interacting via CC interaction produce an associated lepton. The DIS events produce a number of hadrons along with the lepton, while RS interactions produce at most one hadron. In the QE process, no hadrons are produced and the final-state lepton takes away most of the energy of the incident neutrino.
\paragraph*{}
 A \verb!C++! code developed by the collaboration using the GEANT4-based \cite{GEANT4}  simulation toolkit, containing the full ICAL detector geometry, magnetic field map and RPC characteristics, is used to propagate the secondary particles. The signals induced by the events in the RPC are digitalized to form the position $(x,z)$ or $(y,z)$ and time $t$, referred to as hits. The hits are fit to form the tracks, and further details of which are discussed in section \ref{sec3b}. Hence, the output from GEANT4 also contains the information on energy loss and  momentum of the particle all along its path.  This paper is based only on the CC neutrino events, where we consider the muon information alone. The information on the energy and direction of the muons is used to study the sensitivity to atmospheric neutrino oscillation parameters $\theta_{23}$ and $\Delta m^{2}_{32}$ at INO-ICAL and resolving the MH. Including hadron information is beyond the scope of this paper.
 
\section{Analysis Procedure}\label{sec3}
The first step in the procedure is event generation. The generated events are reconstructed in the GEANT4-simulated ICAL and oscillations applied event-by-event after event selection. The oscillated events are binned and used in the $\chi^2$ analysis to determine the oscillation parameters. Each of these procedures are described in detail in the subsections below.

\subsection{Event generation}
In this analysis, NUANCE data for an exposure of 50 kton $\times$ 1000 years is generated, out of which sub-samples corresponding to five years of data are used as the experimentally simulated sample and the remaining 995 years of data are used to construct probability distribution functions (PDF) that are used in the $\chi^2$ fit. Hence the data are uncorrelated with the PDFs that are used to fit the data. This paper is based only on the CC neutrino events with energies less than $50~\mathrm{GeV}$ which corresponds to $98.6\%$ of the sample. The idealized case, where the NUANCE data is folded with detector efficiencies and smeared by the resolution functions obtained from GEANT-based studies of single muons with fixed direction and energy,  has been presented previously \cite{INOmu}. In the earlier analysis, although the data was analysed for an exposure of 5 or 10 years, it was scaled down from the 1000 year sample. Hence, the reconstructed central value was always practically the same as the input value. Here we examine in detail the more realistic case, where the data size and central value are both subject to fluctuations.

\subsection{Event reconstruction}\label{sec3b}
The generated NUANCE data is simulated in GEANT4-based detector environment, and for the first time we have done this analysis using event-by-event reconstruction. The tails of the resolution functions, which have been approximated by single Gaussians and Vavilov \cite{Vavilov} functions in the previous studies \cite{INO_phy}, are also taken in to account in this analysis. A charged particle passing through the detector leaves hits in the RPCs. The mutually orthogonal copper strips on the RPC give the $x$ and $y$ position of the hit in each layer, while the layer number gives the $z$ position. Also the timing information from the RPCs, with a resolution of approximately $1~\mathrm{ns}$ enables the distinction between upward and downward going particles.
 
\begin{figure}[!htb]
\includegraphics[height=0.5\linewidth,width=0.7\linewidth]{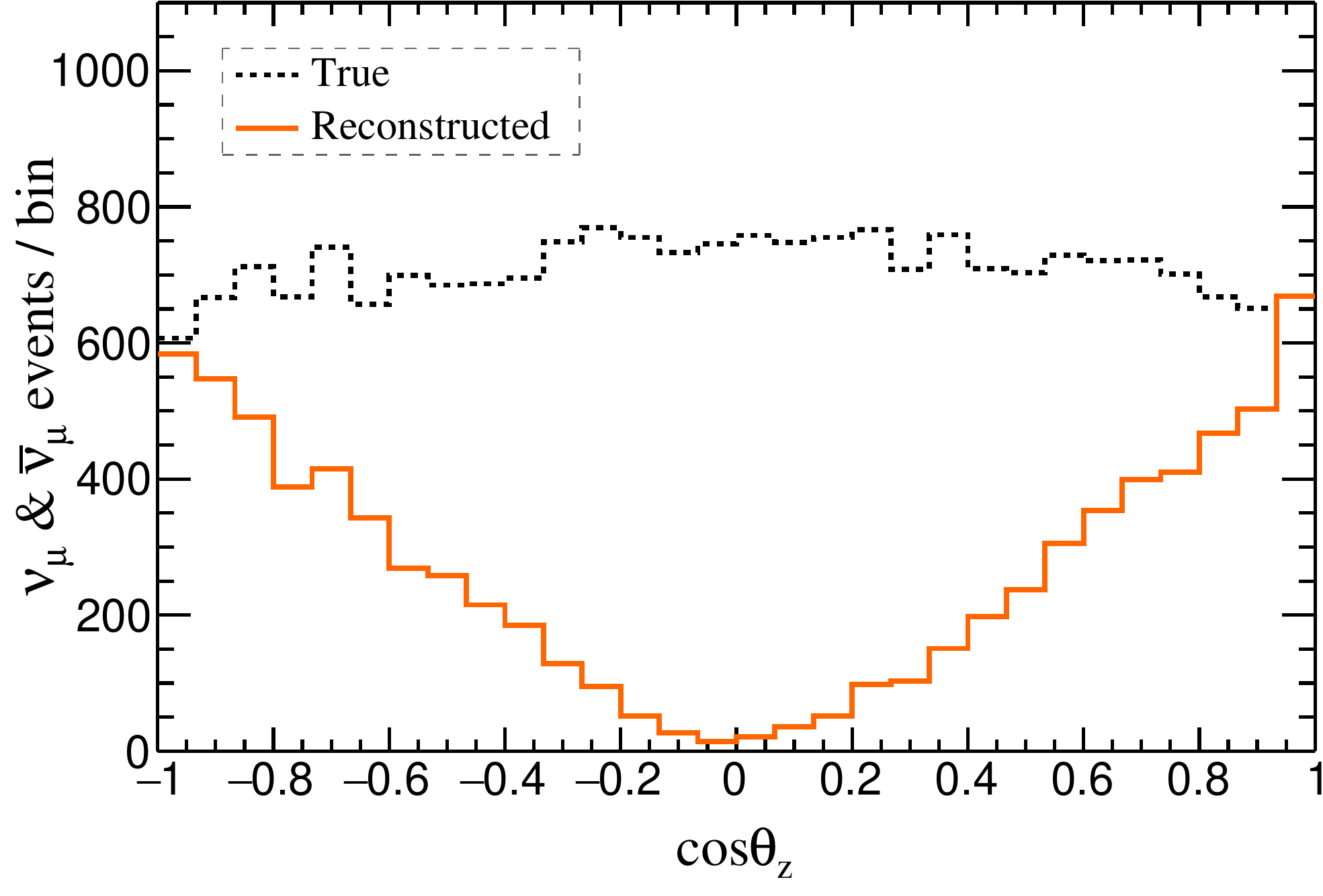}%
\caption{Comparison of true (dashed black) and reconstructed (solid orange) zenith angle ($\cos\theta_z$) distribution for muons, averaged over energy and azimuthal angle for an exposure of $50~ \mathrm{kton} \times 5~\mathrm{years}$ of ICAL.\label{fig0}}
\end{figure}

\paragraph*{}
 The $\mu^\pm$ being minimum ionising charged particles, leave one or two hits per layer on  average, forming a well-defined track, whereas the hadrons leave several hits per layer forming a shower of hits. Rarely (less than 1\% of the time) a pion may also leave a well-defined track in the ICAL and may be misidentified as a muon. In this case the longest track is identified as the muon. The iron plates will be magnetised to produce a field upto $1.5~\mathrm{T}$ and this will be used in the ICAL to probe the charge and momentum of the muon. The direction and the curvature of the muon trajectory, as it propagates through the magnetized detector, gives its charge and momentum, respectively. A recursive optimal state estimator -- the Kalman Filter \cite{Fm, INO_trkfit}, uses  the local geometry and magnetic field information to fit the muon hits, where the muons passing through a minimum of three layers are fit to form the track. The direction and the momentum of the muon is reconstructed from the best fit values of the track. More details can be found in Ref. \cite{INOmu_res}. Figure \ref{fig0} shows the zenith angle $(\theta_z)$ distribution before (true) and after reconstruction. Note that in the current analysis $\cos\theta_{z} = +1$ is the up-going direction.

\paragraph*{} 
The energy of hadrons is obtained by calibrating the number of hits not associated with the muon track, in the event \cite{INOhad_res}. The incident neutrino energy ($E_{\nu}$) can be reconstructed from the energies of the muons and hadrons produced in the detector. The poor energy resolution of hadrons \cite{INOhad_res} affects the reconstruction of the incident neutrino. Hence for ICAL physics analysis hadron and muon energies are used separately, without losing the good energy and angular resolution of muons \cite{INOmu_res,INOmu,INOhad, INO_phy}.

\subsection{Event selection}

The reconstruction of muons is badly affected by the non-uniform magnetic field and dead spaces such as coil slots and support structures. Also the horizontal events which pass through very few layers giving very few hits are reconstructed poorly. Partially contained events, where the $\mu^{\pm}$ leaves the detector volume, are typically harder to reconstruct, as most of the time they leave a short track within the detector. To remove these badly reconstructed events and obtain a better reconstructed sample of data, we investigated applying selection cuts as used in the previous Monte Carlo (MC) studies \cite{INOcutp}.

Events with $\chi^2/\mathrm{ndf}<10$ are used in the analysis, where the $\chi^2$ is the chi-square estimate of the fit for the track obtained from the Kalman filter, and ndf is the number of degrees of freedom. Here $\mathrm{ndf}=2 N_{\mathrm{hits}}-5$, where the Kalman filter fits five parameters to form the track and $N_{\mathrm{hits}}$ are the number of hits associated with the track, with each hit having two degrees of freedom as they are either $(x,z)$ or $(y,z)$ in coordinates. The badly reconstructed horizontal events are removed by applying a  cut of $|\cos\theta_z|\geq 0.35$. Also, to keep a check on the events leaving from top and bottom of the detector, a cut on the $z$-position of the event vertex is applied. Events with vertices lying below $z=6~\mathrm{m}$ and above $z=-6~\mathrm{m}$ are the ones selected from up-going and down-going events respectively.

\begin{figure}[!htb]
  \includegraphics[width=8cm,keepaspectratio]{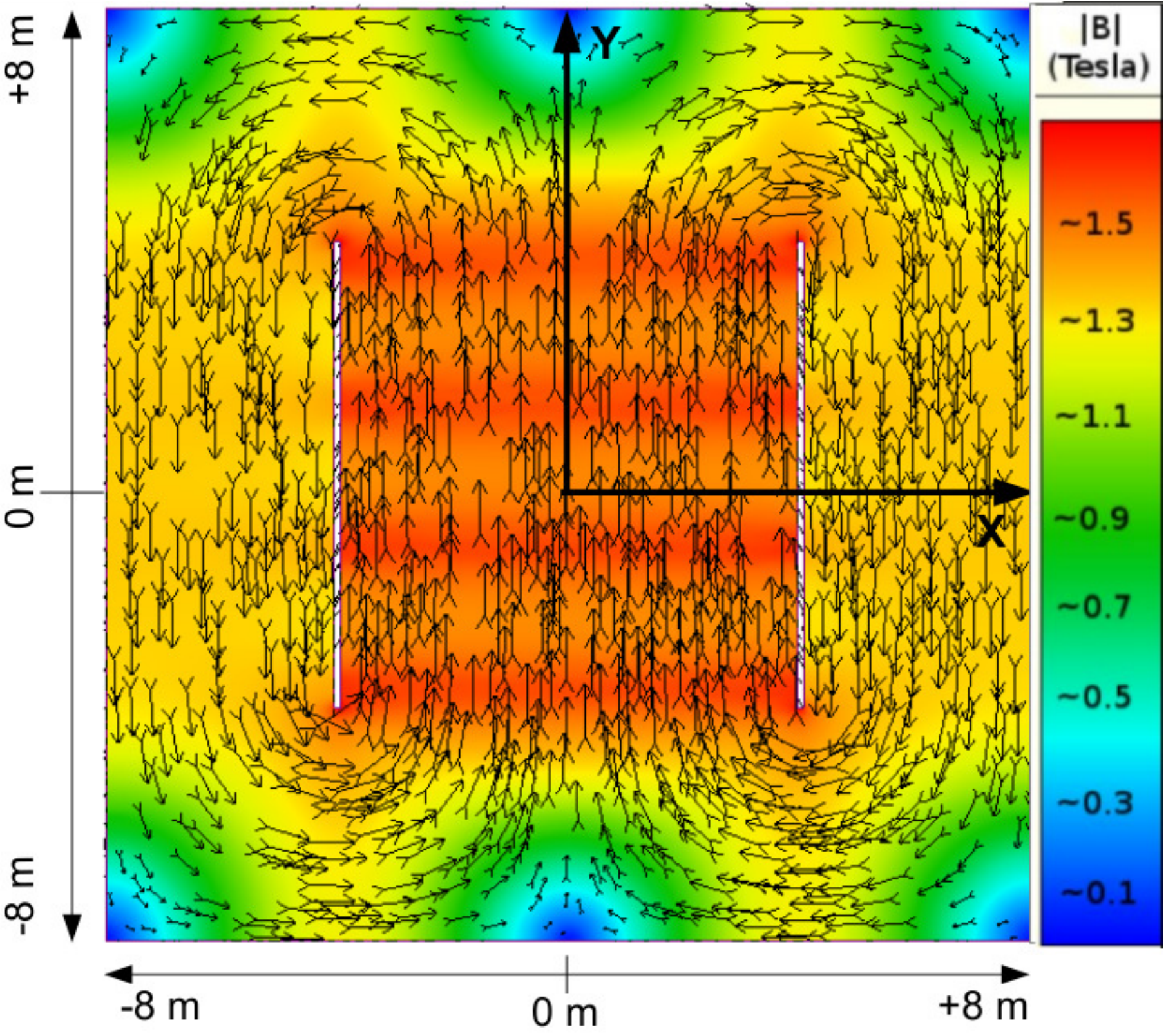}%
\caption{Magnetic field map in the central plate ($z=0$) of the central module \cite{INO_trkfit}. The direction and magnitude (in $\mathrm{T}$) of the magnetic field is shown using the direction and length of arrows respectively. The magnetic field strength is also shown using the color-code. Note that only central module is represented here, but ICAL has three identical modules kept side-by-side\label{fig1}.}
\end{figure}

\paragraph*{}
The entire ICAL detector was divided into three regions, depending on the magnitude of the  magnetic field. Figure \ref{fig1}, shows the magnetic field map in the central iron layer ($z=0$) of the central module (single module). Considering three modules of size $16~\mathrm{m}\times16~\mathrm{m}\times14.4~\mathrm{m}$ each and choosing an origin at the centre of the central module, the ICAL will have conventionally $24~\mathrm{m}$, $8~\mathrm{m}$ and $7.2~\mathrm{m}$ on either side of the origin along $x$, $y$ and $z$ directions respectively. The region $|x|\leq 20~\mathrm{m}$, $|y|\leq 4~\mathrm{m}$, with $z$ unconstrained is defined to be the central region. Here the magnetic field is highest and uniform in magnitude (with $\approx 12\%$ coefficient of variation) despite the fact that the direction of the magnetic field would flip along $y$ in the regions $|x|< 4~\mathrm{m}$, $4~\mathrm{m}\leq|x|< 12~\mathrm{m}$ and $12~\mathrm{m}\leq|x|< 20~\mathrm{m}$. In contrast, the region $|y|> 4~\mathrm{m}$, termed peripheral region, has maximally varying magnetic field in both magnitude (with $\approx 28\%$ coefficient of variation) and direction. Finally the third region $|x|> 20~\mathrm{m}$ and $|y|\leq 4~\mathrm{m}$, termed the side region, has a magnetic field smaller by $\approx 11\%$ and opposite in direction to the central region. The side region has better uniform magnetic field among the three regions (with less than $5\%$ coefficient of variation).

\paragraph*{}
All the events with the interaction vertices in the central region, and with $N_{\mathrm{hits}}>0$, are selected as they are either contained within the detector or can form a reasonable length of track to identify the direction and momentum. The rest of the events in the peripheral and side regions are classified into partially (PC) and fully (FC) contained events according to the end position of the track. If the track end lies within  $|x|\leq 23~\mathrm{m}$ and $|y|\leq 7.5~\mathrm{m}$ and $z\leq 7~\mathrm{m}$, then the event is classified as FC and is selected. The remaining events are classified as PC, and a selection criterion of $N_{\mathrm{hits}}>15$ is applied on all such PC events. Table \ref{tab2} summarise the selection criterion used in this analysis.

\begin{table}[!htb]
\caption{\label{tab2}Lists the region wise and the overall (common) selection criterion used in this current analysis}.
\begin{ruledtabular}
\begin{tabular}{cccc}
Region & Event type & Region specific selection & Common selection \\
\hline
\multirow{1}{*}{Central}& all & $\mathrm{nhits}>0$ &\multirow{2}{*}{$\chi^2/\mathrm{ndf}<10$} \\
\multirow{2}{*}{Side}& FC & $\mathrm{nhits}>0$ & \\
&PC & $\mathrm{nhits}>15$ & \multirow{1}{*}{ $|\cos\theta_z|\geq 0.35$}\\
\multirow{2}{*}{Peripheral}& FC & $\mathrm{nhits}>0$ & \multirow{1}{*}{$z_{\rm vertex} < 6~\mathrm{m}$ (up going)}\\
&PC & $\mathrm{nhits}>15$ & \multirow{1}{*}{$z_{\rm vertex} >-6~\mathrm{m}$ (down going)}\\
\end{tabular}
\end{ruledtabular}
\end{table}

\begin{figure}[!htb]
\subfloat[\label{fig2a}]{%
\raisebox{0.014\textwidth}{%
  \includegraphics[height=0.4\linewidth,width=0.49\linewidth]{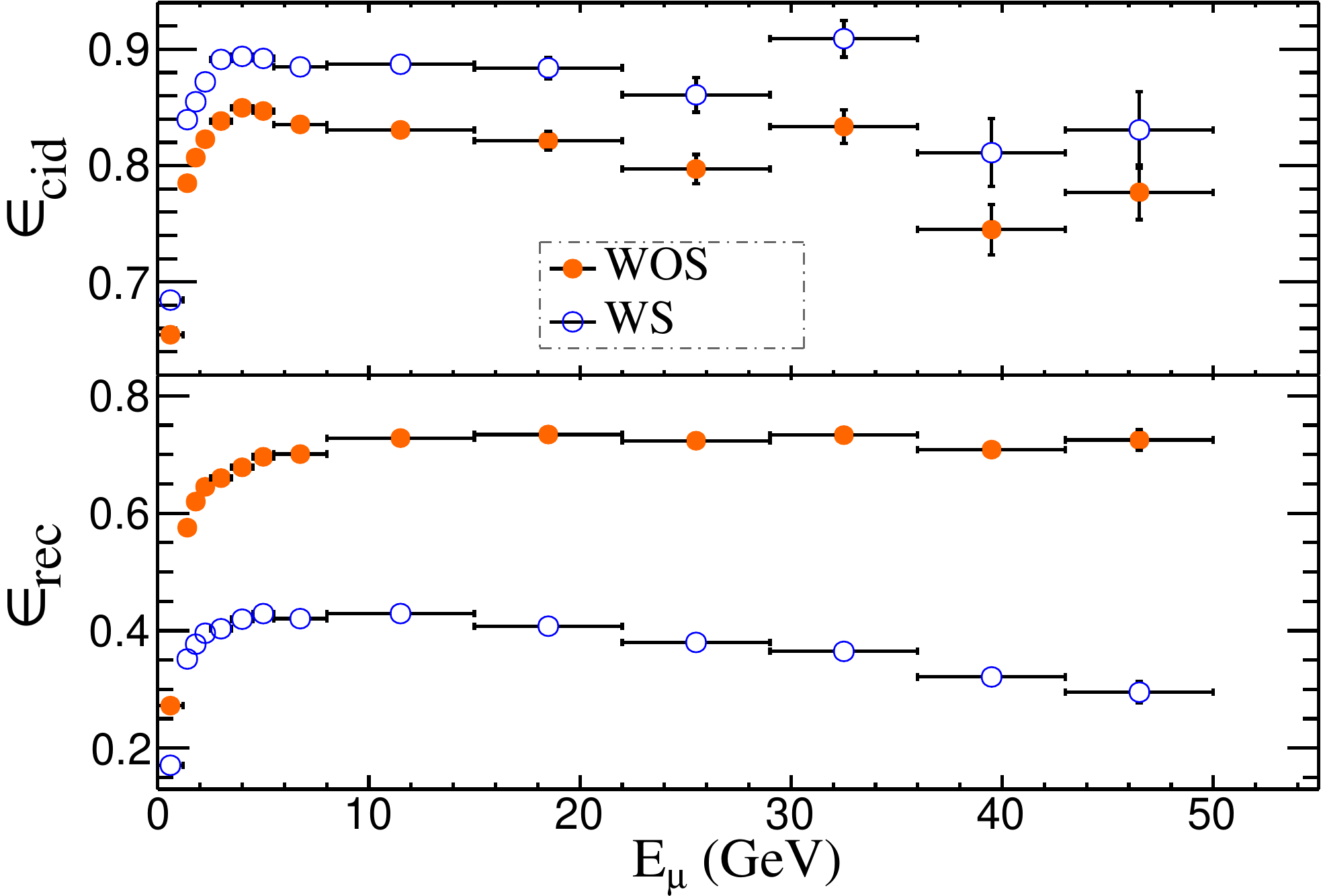}%
  }
}\hfill
\subfloat[\label{fig2b}]{%
  \includegraphics[height=0.41\linewidth,width=0.49\linewidth]{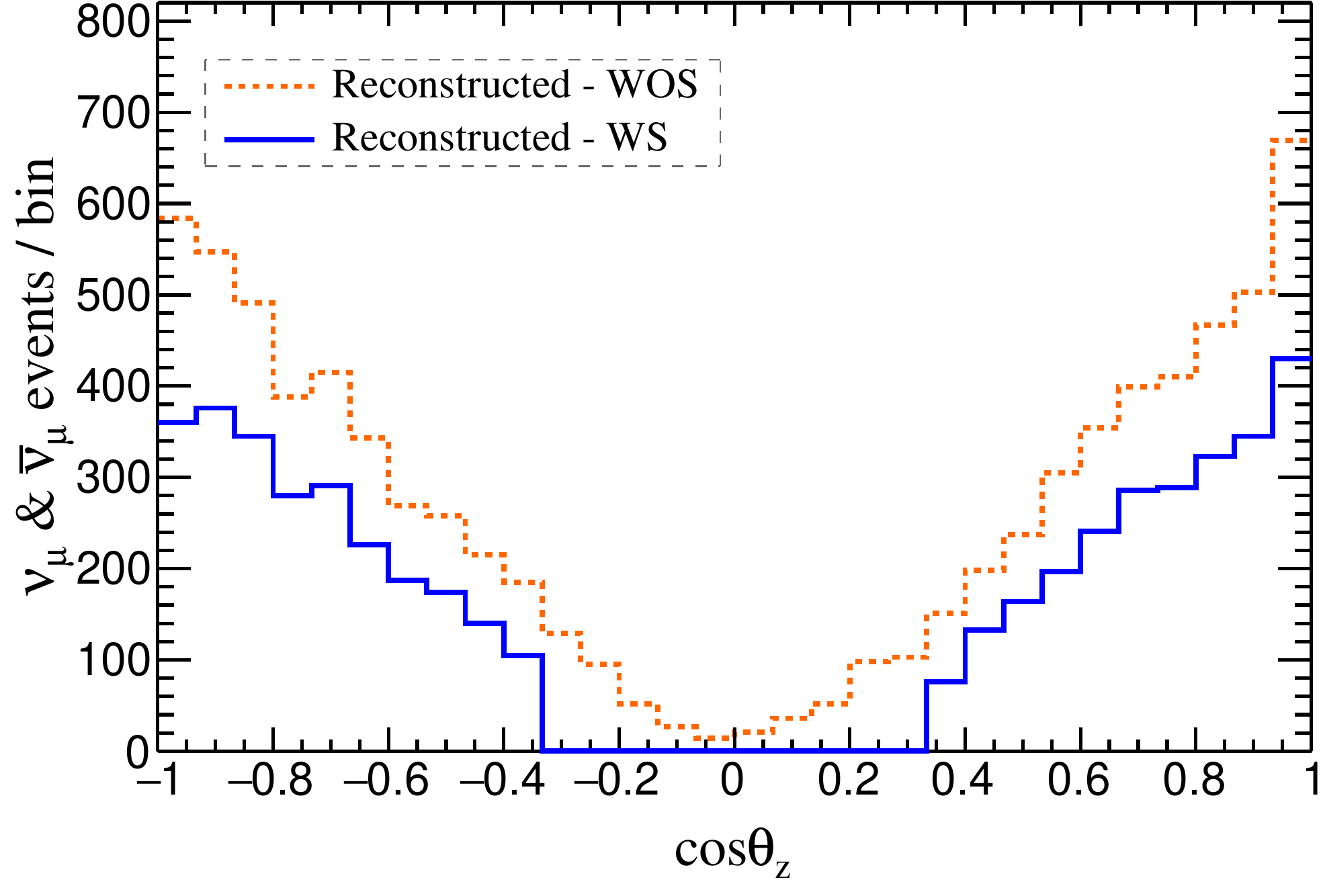}%
}
\caption{\label{fig2}\protect\subref{fig2a} Reconstruction efficiency $\epsilon_{\mathrm{rec}}$ (bottom panel) and CID efficiency $\epsilon_{\mathrm{cid}}$ (top panel) for the events with (WS) (open circles) and without (WOS) (solid circles) event selection, as a function of true muon energy $E_{\mu}$. The statistical error corresponds to $50~\mathrm{kton}\times 100~\mathrm{years}$ reconstructed data sample. \protect\subref{fig2b}~Comparison of reconstructed zenith angle ($\cos\theta_z$) for WOS (dashed orange) and WS (solid blue), averaged over energy and azimuthal angle for an exposure of $50~ \mathrm{kton} \times 5~\mathrm{years}$ of ICAL.}

\end{figure}
The reconstruction efficiency ($\epsilon_{\mathrm{rec}}$) \ie the fraction of events reconstructed from the total number of events, is shown in Fig. \subref*{fig2a} as a function of true muon energy. For $E_{\mu}<6~\mathrm{GeV}$, the reconstruction efficiency increases with increase in muon energy, as the energetic muon pass through large number of layers. At higher energies the reconstruction efficiencies become almost a constant. The relative charge ID (CID) efficiency ($\epsilon_{\mathrm{cid}}$) \ie the fraction of events identified with correct muon charge among the total reconstructed events, is compared with (WS) and without (WOS) event selection in Fig. \subref*{fig2a}. The CID efficiency increases after applying selection cuts, as most of the badly reconstructed events are discarded.
  
  Approximately $40\%$ of reconstructed events are lost with event selection. Figure \subref*{fig2b}, shows the reconstructed $\cos\theta_z$ distribution and compares the reconstructed events with (WS) and without selection (WOS) criterion. The dip at $\cos\theta_{z}=0$ results from the difficulty to reconstruct horizontal events. This paper also studies the effect of the applied event selection on the sensitivity of oscillation parameters in ICAL. Hence, the parameter sensitivity is studied with and without applying any selection criterion.

\subsection{Applying Oscillations}
\paragraph*{}
The muon signal in the ICAL will have contributions from the component of the $\nu_{e}$ flux $(\Phi_{\nu_e})$ that has oscillated to $\nu_{\mu}$ and the component from $\nu_{\mu}$ flux $(\Phi_{\nu_{\mu}})$ that has survived. As the five-year pseudo-data would have negligible contribution from $\Phi_{\nu_e}$ compared to $\Phi_{\nu_{\mu}}$, we have only used the $\Phi_{\nu_{\mu}}$ events in our analysis. Hence, neglecting oscillations from $\Phi_{\nu_e}$, the total number of events appearing in the detector for an exposure time $T$ is obtained from
\begin{equation}\label{eq1}
\frac{d^2N}{dE_{\nu} \  d\cos\theta_{z}}=T \times N_D \times \sigma_{\nu_{\mu}} \left[P_{\mu\mu}\frac{d^2\Phi_{\nu_{\mu}}}{dE_{\nu} \  d\cos\theta_{z}}\right],
\end{equation}
where $N_D$ is the number of targets in the detector and $P_{\mu\mu}$ is the $\nu_\mu$ survival probability. Oscillation probabilities are calculated by numerically evolving the neutrino flavor eigenstates \cite{Indu_prem} using the equation,
\begin{equation}\label{eq2}
i\frac{d}{dt}\left[\nu_{\alpha}\right]=\frac{1}{2E}\left(UM^2U^{\dagger}+\mathcal{A}\right)\left[\nu_{\alpha}\right],
\end{equation}
where  $\left[\nu_{\alpha}\right]$ denotes the vector of flavor eigenstates, $\nu_{\alpha}$, $\alpha = e, \mu, \tau$, $U$ is the PMNS mixing matrix, and $M^2$ is the mass squared matrix. Here, $\mathcal{A}$ is the diagonal matrix, diag$(A,0,0)$, with matter term $A$ given by

\vspace*{-0.5cm}
\begin{align}\label{eq3}
\begin{split}
A  = & ~\pm 2\sqrt{2}G_Fn_eE \\
=& ~\pm 7.63 \times 10^{-5}~\rho E,
\end{split}
\end{align}
where the sign is positive for $\nu$ and negative for $\bar\nu$. Here, $G_F$ is the Fermi coupling constant, $E$ is the neutrino energy in $\mathrm{GeV}$, and $n_e$ is the electron number density which is related to the matter density $\rho$ in $\mathrm{gcm^{-3}}$. The density profile of the Earth's matter, given by Preliminary Reference Earth Model (PREM) \cite{prem}, is used to calculate oscillation probabilities for $\nu$ and $\bar\nu$. The difference in sign of $A$ for $\nu$ and $\bar\nu$ leads to differing oscillation probabilities, which in turn are sensitive to the sign of $\Delta m^2_{32}$. The ICAL has an advantage because it can differentiate between $\nu$ and $\bar\nu$ events and observe the matter effects separately.

\begin{figure}[!htb]
\subfloat[\label{fig3a}]{%
  \includegraphics[height=0.38\linewidth,width=0.495\linewidth]{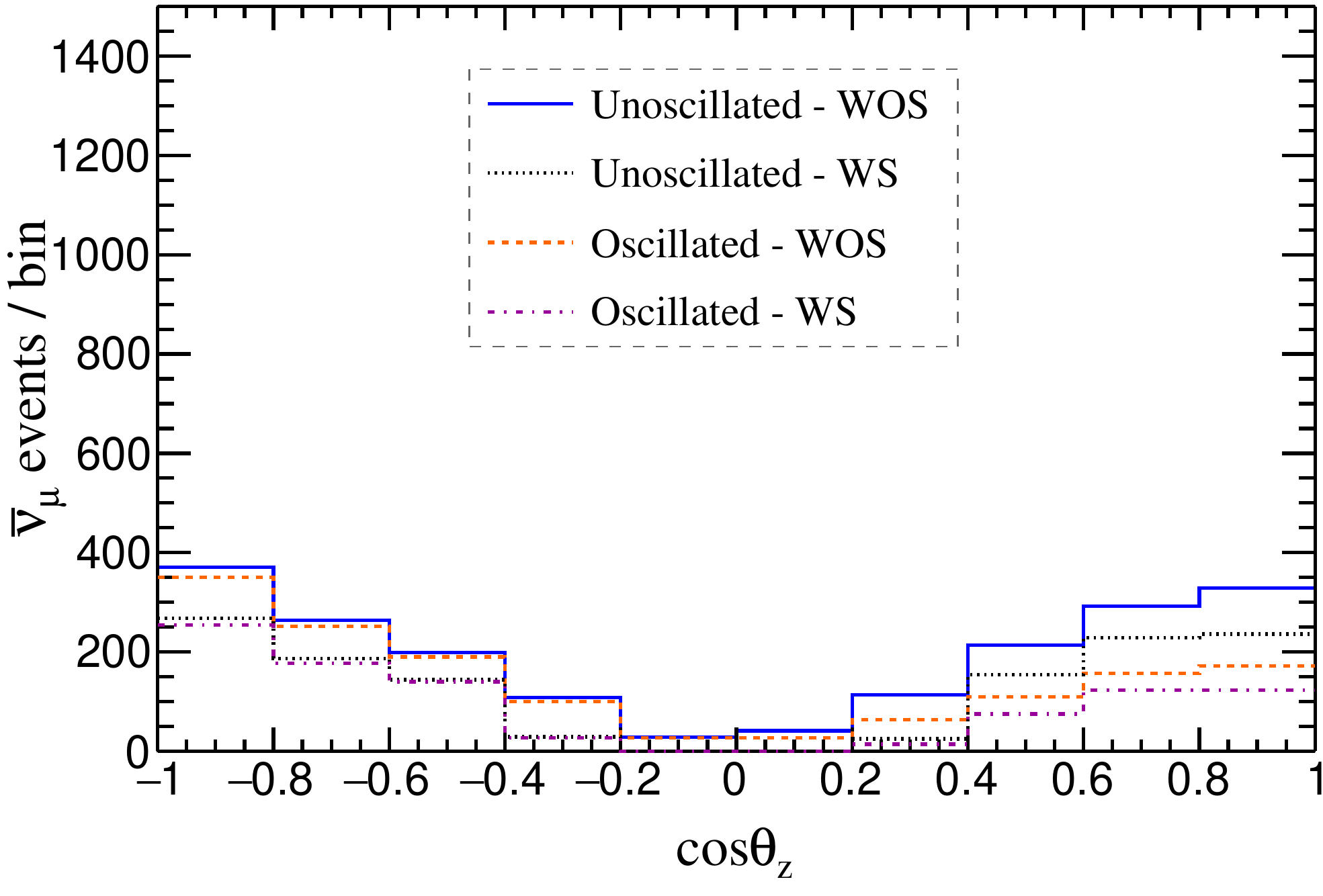}%
}\hfill
\subfloat[\label{fig3b}]{%
  \includegraphics[height=0.38\linewidth,width=0.495\linewidth]{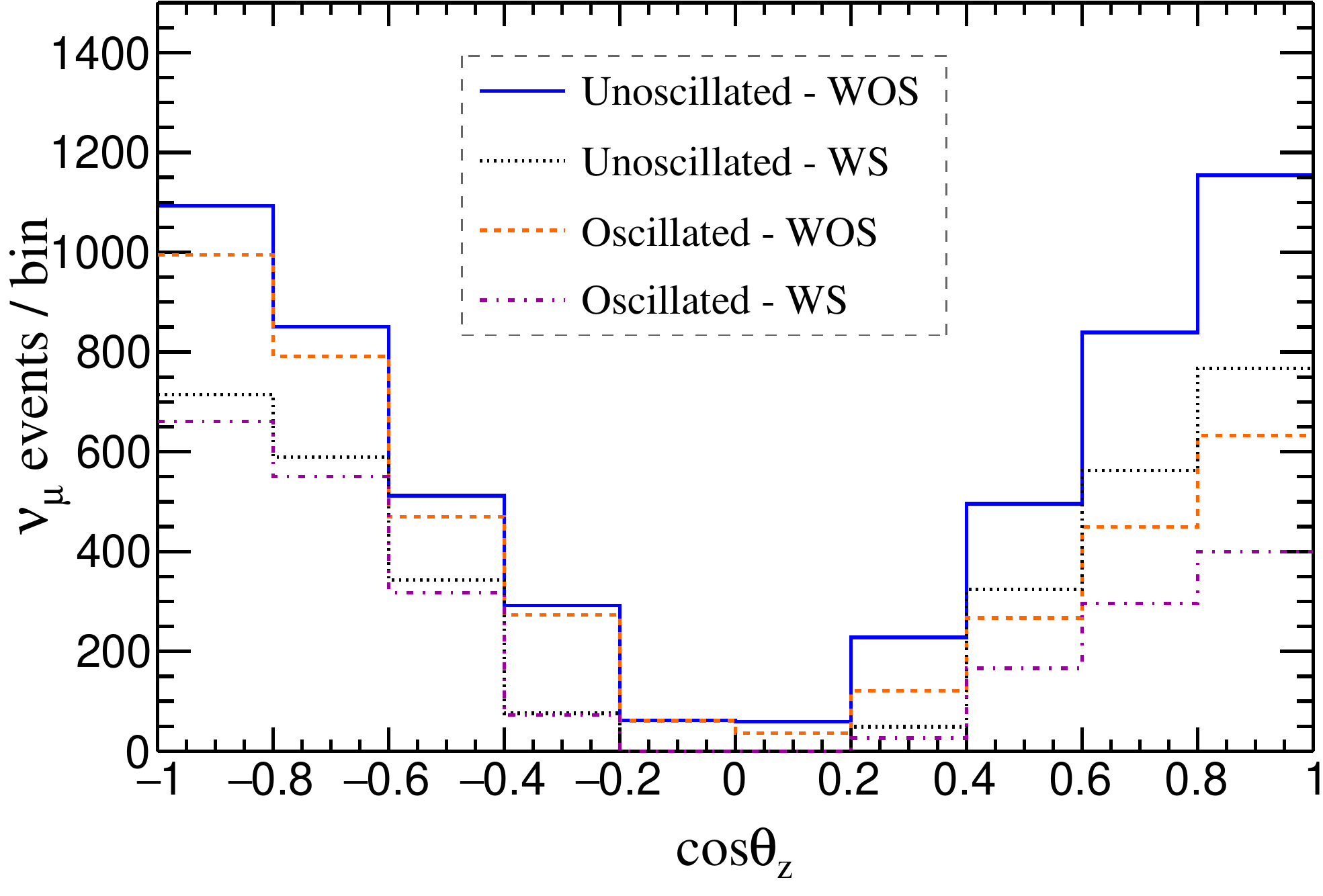}%
}
\caption{\label{fig3}$\cos\theta_z$ distributions with and without oscillations for \protect\subref{fig3a} $\mu^+$ obtained from $\bar\nu_{\mu}$ events. \protect\subref{fig3b} $\mu^-$ obtained from $\nu_{\mu}$ events. The distributions are also compared with (WS) and without (WOS) selection criterion.}

\end{figure}

\paragraph*{}
Oscillations are applied on the five-year data sample using the accept or reject method. First, the survival probability $P_{\mu\mu}$ is calculated for each $\nu$ or $\bar\nu$ with a given energy and direction. To decide whether an un-oscillated $\nu_{\mu}$ survives oscillations to be detected as $\nu_{\mu}$, a uniform random number $r$ is generated between 0 and 1. If $P_{\mu\mu}>r$, the event is accepted to have survived the oscillations. Otherwise it is considered to have oscillated into another flavor and is rejected. Here, we have used the true values of the oscillation parameters assuming NH, from Ref. \cite{PDG_14}; see Table \ref{tab3}. The zenith angle distribution of muons before and after applying oscillations via the accept or reject method is shown in Fig. \subref*{fig3a} and \subref*{fig3b} for $\bar\nu_{\mu}$ and $\nu_{\mu}$ events respectively. It also compares the zenith angle distributions with (WS) and without (WOS) event selection. The reduction in upward going events $(\cos\theta_z>0)$ is evident, as the $\nu$ or $\bar\nu$ would travel a much larger distance compared to downward going neutrinos $(\cos\theta_z<0)$, thus having a larger probability to oscillate into another flavor. Also, the oscillation signatures are different in $\bar\nu_{\mu}$ and $\nu_{\mu}$ events, where it depends on the sign of $\Delta m^2_{32}$. This difference is solely due to the matter effects, as we have assumed no $CP$ violation. (It has been clearly established that CC $\mu$ events in the ICAL are insensitive to $\delta_{\rm cp}$ \cite{INO_phy}.) Hence $\nu_{\mu}$ events are separated from $\bar\nu_{\mu}$ events while binning, to have maximum sensitivity to the MH.

\begin{table}[!htb]
\caption{\label{tab3}Assumed values of oscillation parameters \cite{PDG_14} used to construct the pseudo-data}
\begin{ruledtabular}
\begin{tabular}{lc}
Parameter & Input Value\\
\hline
 $\sin^2\theta_{23}$ & 0.5\\
 $\sin^2\theta_{12}$ & 0.304\\
 $\sin^2\theta_{13}$ & 0.0219\\
 $\Delta m^2_{21}$ $(\mathrm{eV^2})$ & 7.53$\times$10$^{-5}$\\
 $\Delta m^2_{32}$ $(\mathrm{eV^2})$ & 2.32$\times$10$^{-3}$\\
 $\delta_{cp}$\footnote{$\delta_{cp}$ is assumed to be zero} & 0\\
\end{tabular}
\end{ruledtabular}
\end{table}

\subsection{Binning scheme}
\paragraph*{}
During reconstruction, the positively charged particles are tagged with positive momentum and the negatively charged particles are tagged with negative momentum by convention. Muons with positive reconstructed momentum are therefore identified as $\mu^+$ from an anti-neutrino event, and the ones with negative momentum are identified as $\mu^-$ from a neutrino event.  The reconstructed  muons of positive and negative charges are binned separately in $Q_{\mu}E_{\mu}$ and $\cos\theta_z$ bins after applying oscillations, where $Q_{\mu}=\pm1$ for $\mu^+/\mu^-$. The events with negative and positive $Q_{\mu}E_{\mu}$ indicate those identified as $\nu_{\mu}$ and $\bar\nu_{\mu}$ events respectively, based upon the charge identification from the curvature of the reconstructed track. The atmospheric neutrino flux falls rapidly at higher energies. Hence, wider bins were chosen in those energy regions to ensure adequate statistics. Also, within the frame work of low event statistics, increasing the number of high energy bins is not feasible due to the limited statistics. Table \ref{tab4} summarises the binning scheme used in the current analysis. 
\paragraph*{}
The effect of finer binning is studied previously \cite{INOmu} for energies less than $11~\mathrm{GeV}$, and is known to marginally improve the precision in both $\sin^2\theta_{23}$ and $|\Delta m^2_{32}|$. Also, increasing the range of energies beyond $11~\mathrm{GeV}$ is known to improve the result \cite{Ino_lak}. Increasing the number of high energy bins can improve the precision. The optimization of bin widths at higher energies will be a part of the future work, and the current analysis will focus on the effects of fluctuations arising from the low event statistics.

\begin{table}[!htb]
\caption{\label{tab4}The binning scheme for the reconstructed observables $\cos\theta_z$ and $E_{\mu}$}.
\begin{ruledtabular}
\begin{tabular}{ccccc}
Observable & Range & Bin width & Bins & Total bins \\
\hline
\multirow{6}{*}{$E_{\mu}$ $(\mathrm{GeV})$}& [-1.2, -0.2], [0.2, 1.2] & 1.0 & 2 &\multirow{6}{*}{18}\\
& [-2, -1.2], [1.2, 2] & 0.4 & 4 &\\
& [-2.5, -2], [2, 2.5] & 0.5 & 2 &\\
& [-5.5, -2.5], [2.5, 5.5] & 1.0 & 6 &\\
& [-8, -5.5], [5.5, 8] & 2.5 & 2 &\\
& [-50, -8], [8, 50] & 42 & 2 &\\
\hline
$\cos\theta_z$ & [-1, 1] & 0.2 & 10 & 10\\
\end{tabular}
\end{ruledtabular}
\end{table}

\subsection{The $\chi^2$ analysis and systematics}\label{3d}
\begin{figure}[!htb]
\subfloat[\label{fig5a}]{%
  \includegraphics[height=0.38\linewidth,width=0.495\linewidth]{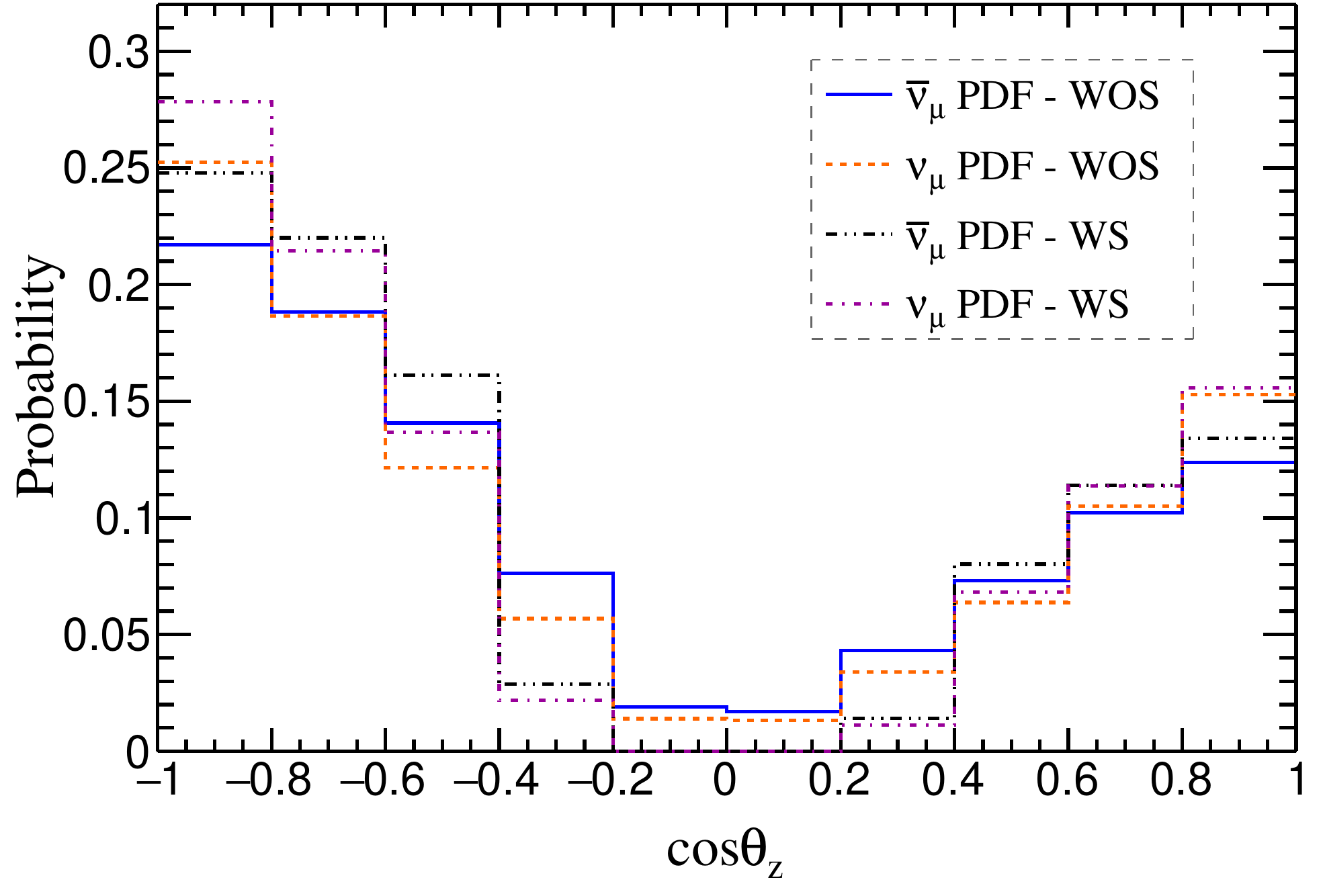}%
}\hfill
\subfloat[\label{fig5b}]{%
  \includegraphics[height=0.38\linewidth,width=0.495\linewidth]{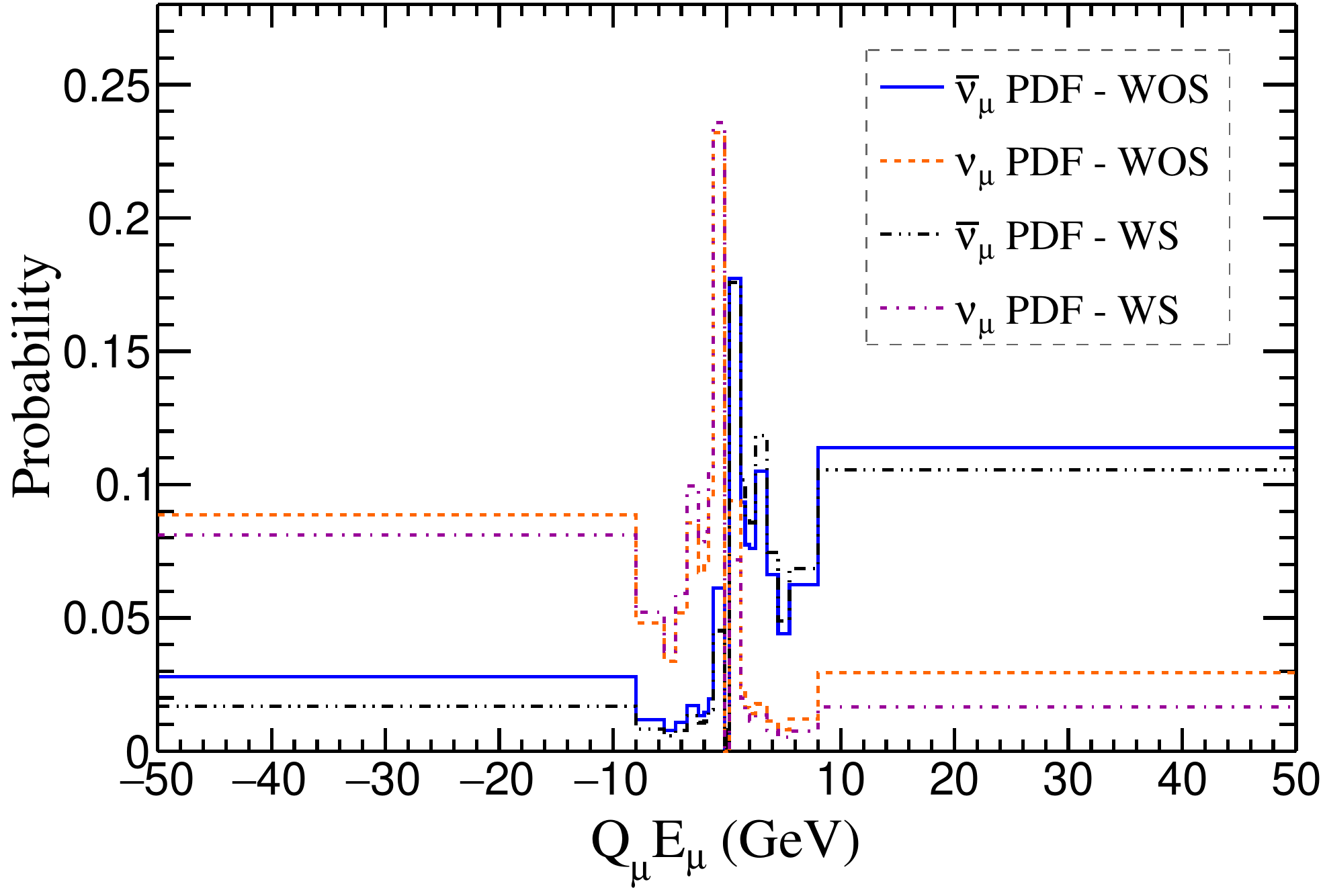}%
}
\caption{\label{fig5}PDF's for $\bar\nu_\mu$ and $\nu_\mu$ are shown for \protect\subref{fig5a} binning in $\cos\theta_z$, and \protect\subref{fig5b} binning in $E_{\mu}$. The $\bar\nu_{\mu}~(\nu_{\mu})$ entries for $q_{\mu}E_{\mu}<0~(>0)$ indicate the charge mis-id content.}

\end{figure}

The pull approach \cite{pull} is used in defining the $\chi^2$ such that systematic uncertainties are incorporated. The pull approach is equivalent to the covariance approach, but is computationally much faster. After binning the oscillated events, 
the five year simulated data set is fit by defining the following $\chi^2$ \cite{sk_sys,INOmu}:
\begin{widetext}
\begin{equation}\label{eq4}
\chi^2 = \min_{\{\xi_k\}}\sum_{i=1}^{n_{\cos\theta_{z}}}\sum_{j=1}^{n_{E_{\mu}}}2\left[\left(N_{ij}^{\mathrm{pdf}}-N_{ij}^{\mathrm{data}}\right)-N_{ij}^{\mathrm{data}}\ln\left(\frac{N_{ij}^{\mathrm{pdf}}}{N_{ij}^{\mathrm{data}}}\right)\right] + \sum_{k=1}^{2}\xi_k^2,
\end{equation}
\end{widetext}
where,
\begin{equation}\label{eq5}
N_{ij}^{\mathrm{pdf}}= R\left[f T_{ij}^{\bar\nu} +(1-f) T_{ij}^{\nu}\right]\left[1+\sum_{k=1}^{2}\pi_{ij}^k\xi_k\right].
\end{equation}
Here, $N_{ij}^{\mathrm{data}}$ and $N_{ij}^{\mathrm{pdf}}$ are the observed and the expected number of muon events respectively, in a given ($ \cos\theta_{z}^i,E_{\mu}^j$) bin, while $n_{\cos\theta_{z}}$ and $n_{E_{\mu}}$ are the total number of $\cos\theta_{z}$ and $E_{\mu}$ bins respectively. $N_{ij}^{\mathrm{data}}$ is calculated for true values of oscillation parameters, summarised in Table \ref{tab1}, whereas $N_{ij}^{\mathrm{pdf}}$ is obtained by combining $\nu_\mu$ and $\bar\nu_\mu$ PDFs as in Eq. \ref{eq5}, where $T_{ij}^{\nu}$ and $T_{ij}^{\bar\nu}$ are the $\nu$ and $\bar\nu$ PDFs respectively normalized from 995 year sample, with $f$ as a free parameter describing the relative fraction of $\bar\nu_\mu$ and $\nu_{\mu}$ in the sample, with $R$ being the normalization factor in the fit which scales the PDF to 5 years. The $\bar\nu_\mu$ and $\nu_\mu$ PDFs with (WS) and without (WOS) selection criterion are shown in Figs \subref*{fig5a} and \subref*{fig5b}.
  
\paragraph*{}
The systematic errors and the theoretical uncertainties are parametrized in terms of variables $\left\lbrace \xi_k \right\rbrace$ called pulls. The value $\xi_k = 0$ corresponds to the expected value, and the variation, $\xi_k = \pm1$ corresponds to a one standard deviation for each source of systematics resulting in an uncertainty of  $\pi_{ij}^k$ for the $k^{\mathrm{th}}$ source.

In this analysis we have considered two systematic uncertainties, a 5\% uncertainty on the zenith angle dependence of the flux and another 5\% on the energy dependent tilt error, parametrized by $\xi_{\mathrm{zenith}}^{\mathrm{flux}}$ and $\xi_{\mathrm{tilt}}^{\mathrm{flux}}$ respectively. There is no systematic uncertainty related to the flux normalization as $R$ and $f$ are the fit parameters which fixes the overall and  relative flux normalizations. To calculate the energy tilt error \ie the possible deviation of the energy dependence of the atmospheric fluxes from the power law, we use the standard procedure as given in, for example Ref \cite{sk_sys}, and define:

\begin{equation}\label{eq6}
\Phi_{\delta}(E)= \Phi_{0}(E)(\frac{E}{E_0})^{\delta} \approx \Phi_{0}(E)\left[1+\delta \ln\frac{E}{E_0}\right],
\end{equation}
Neglecting the effect of oscillations, the expected number of events $\Phi_{0}(E)$ is calculated for each $(ij)^{th}$ bin. Then we compute $\Phi_{\delta}(E)$, where $\delta$ is the $1\sigma$ tilt error taken to be 5\% and $E_0$ = 2 $\mathrm{Gev}$, and find the relative change in flux to obtain the coupling $\pi_{ij}^{\rm tilt}$. The coupling $\pi_{ij}^{\rm zenith}$ in each bin is calculated in proportion to the zenith angle value of that particular bin. The parameters in the fit are marginalized as given in Table \ref{tab5}, where $f$ and $R$ are always marginalized over the given ranges. The parameters $\sin^2\theta_{13}$, $\sin^2\theta_{12}$ and $\Delta m^2_{21}$ had minimal effect when marginalized hence were kept constant in the fit without any prior constraint.
\begin{table}[!htb]
\caption{\label{tab5}Marginalization of the parameters as used in the fit}
\begin{ruledtabular}
\begin{tabular}{lc}
Parameter & Marginalization range\\
\hline
 $\sin^2\theta_{23}$\footnote{Marginalized when the data is fit to determine $\Delta m^2_{32}$} & [0,1]\\
 $\Delta m^2_{32}$ $(\mathrm{eV^2})$\footnote{Marginalized when the data is fit to determine $\sin^2\theta_{23}$} & [0.0005,0.005]\\
 $f$ & [0,1]\\
 $R$ & Unconstrained\\
 $\sin^2\theta_{13}$ & Not marginalized\\
 $\sin^2\theta_{12}$ & Not marginalized\\
 $\Delta m^2_{21}$ $(\mathrm{eV^2})$ & Not marginalized\\
 $\delta_{cp}$& Not marginalized\\
\end{tabular}
\end{ruledtabular}
\end{table}

\section{Parameter Determination}\label{sec4}

\begin{figure}[!htb]
\subfloat[\label{fig6a}]{%
  \includegraphics[height=0.38\linewidth,width=0.495\linewidth]{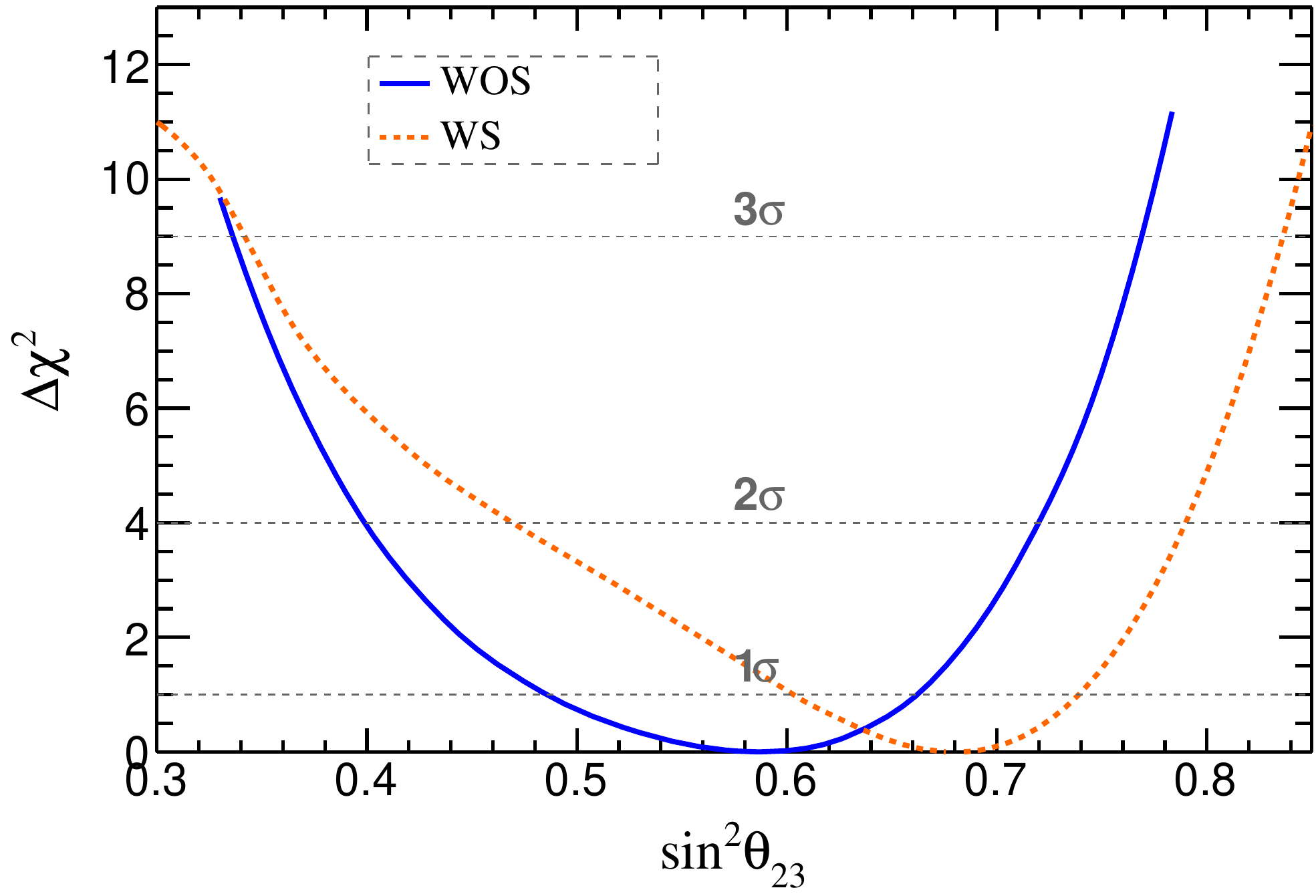}%
}\hfill
\subfloat[\label{fig6b}]{%
  \includegraphics[height=0.38\linewidth,width=0.495\linewidth]{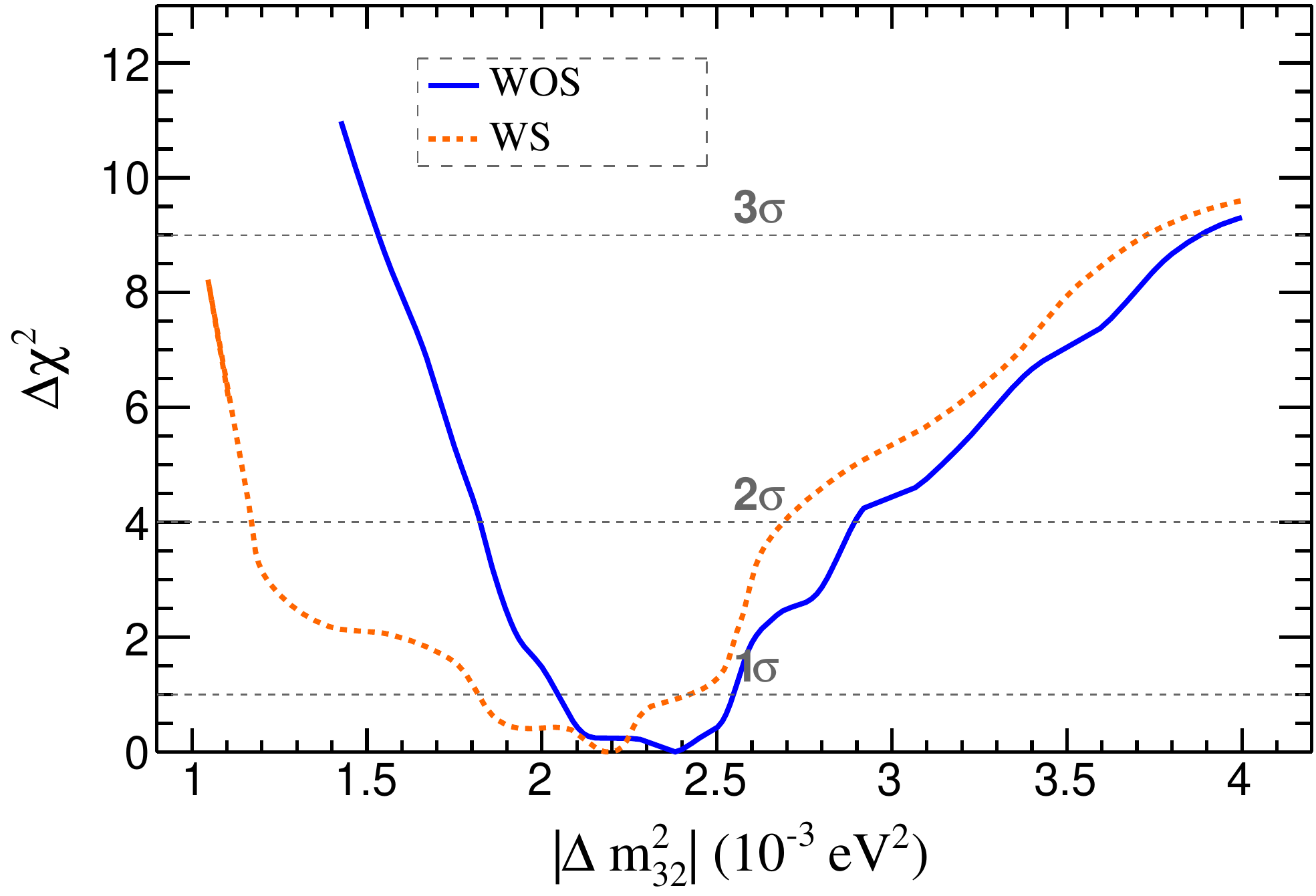}%
}
\caption{\label{fig6}$\Delta\chi^2$ as a function of \protect\subref{fig6a}$\sin^2\theta_{23}$, for an input value of $\sin^2\theta_{23}(\hbox{true}) = 0.5$ and \protect\subref{fig6b} $\Delta m^{2}_{32}$, for an input value of $\Delta m^{2}_{32}(\hbox{true})= 2.32\times 10^{-3}~\mathrm{eV^2}$. The dashed (orange) and solid (blue) line shows the fit with (WS) and (WOS) event selection.}
\end{figure}

The fluctuated pseudo-data set is first fit to determine $\sin^2\theta_{23}$, marginalizing over $|\Delta m^2_{32}|$, for an input value of $\sin^2\theta_{23} = 0.5$. The comparison of $\Delta\chi^2$ with (WS) and without (WOS) event selection is shown as a function of $\sin^2\theta_{23}$ in Fig. \subref*{fig6a}, where the octant degeneracy in  $\sin^2\theta_{23}$, which stems from the leading term $\sin^22\theta_{23}$ in the oscillation probability, is broken due to the relatively large value of $\theta_{13}=8.5\degree$. Hence, the asymmetrical curve in $\sin^2\theta_{23}$ shows the effect of matter oscillations in breaking octant degeneracy. The significance of the fit, \ie how far the observed value (best fit value) is away from the parameters true value (input value), is defined as
\begin{equation}
\mathrm{significance} = \sqrt{\Delta\chi^2_{\mathrm{input}}-\Delta\chi^2_{\mathrm{min}}}~,
\end{equation}
where $\Delta\chi^2_{\mathrm{input}}$ and $\Delta\chi^2_{\mathrm{min}}$ are the $\Delta\chi^2$ values at the true and observed values of the parameter respectively. The fit to $\sin^2\theta_{23}$ without event selection converges to a value of $0.586^{+0.060}_{-0.093}$ with a significance of $0.86$, \ie within $1\sigma$ of the input value, whereas the fit after event  selection converges to $0.676^{+0.063}_{-0.072}$ within $2\sigma$ of the input value. The fit to $\sin^2\theta_{23}$ with event selection shows relatively larger uncertainty at $2$ and $3\sigma$ range.   

The comparison of $\Delta\chi^2$ with and without event selection is shown as a function of $|\Delta m^2_{32}|$ in Fig. \subref*{fig6b}, where the data is fit to determine $|\Delta m^2_{32}|$, marginalizing over $\sin^2\theta_{23}$, for an input value of $\Delta m^2_{32} = 2.32\times 10^{-3}~\mathrm{eV^2}$. The fit without event selection converges to a value of $(2.38^{+0.11}_{-0.39})\times 10^{-3}~\mathrm{eV^2}$, within $1\sigma$ of the input value with a significance of $0.51$, whereas the fit after event  selection converges to $(2.184^{+0.23}_{-0.37})\times10^{-3}~\mathrm{eV^2}$ also within $1\sigma$ of the input value. The fit to $|\Delta m^2_{32}|$ also shows relatively larger uncertainty at $2$ and $3\sigma$ range after applying event selection. The multiple local minimas in $\Delta\chi^2$ function is due to the statistical uncertainty on the PDF, and it is observed to reduce with fits to PDFs constructed from larger MC samples.

\begin{figure}[!htb]
\includegraphics[height=0.5\linewidth,width=0.7\linewidth]{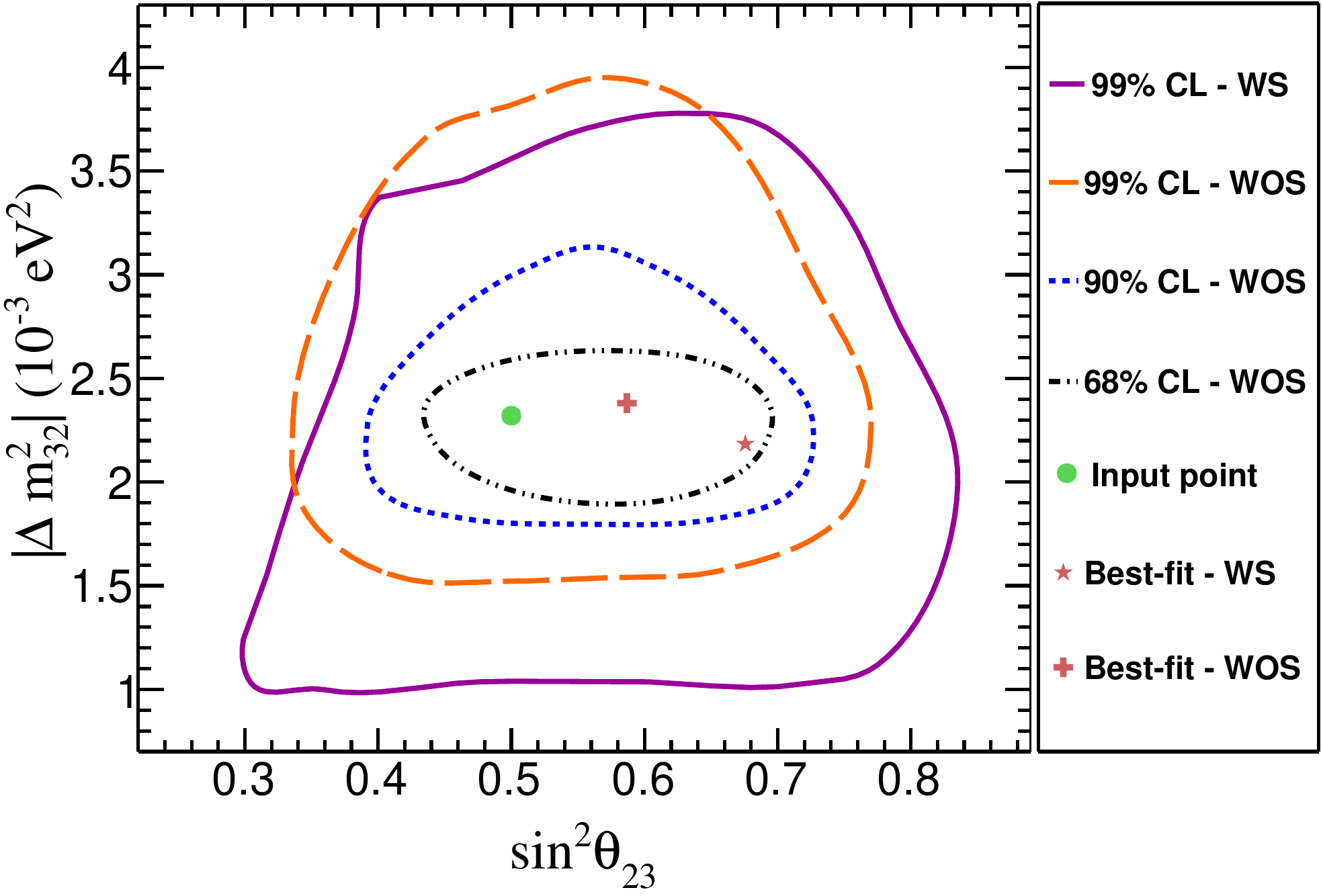}%
\caption{Precision reach obtained from the fit to five year pseudo-data in $\sin^2\theta_{23}-\Delta m^2_{32}$ plane. The dashed (orange), dotted (blue) and the broken (black) line shows the coverage area with $99\%$, $90\%$ and $68\%$ CL respectively without (WOS) event selection, whereas the solid (magenta) line shows the coverage area with $99\%$ CL with (WS) event selection. The input (true) point is given by the green dot and the plus sign signifies the best fit (converged) point.\label{fig7}}
\end{figure}

The parameters $\sin^2\theta_{23}$ and $\Delta m^2_{32}$ are correlated; Fig. \ref{fig7} compares the correlated precision reach obtained fitting a five year pseudo-data set with and without event selection. The best-fit point to the fit without event selection is obtained within a significance of $1\sigma$ from the input value, whereas the fit with event selection converges within a significance of $2\sigma$. The best-fit values of the parameters along with the $1\sigma$ asymmetrical errors are given in the Table \ref{tab6}. Further adding prior constraints on $\sin^2\theta_{13}$ and $\Delta m^2_{12}$, was observed not to make any difference in the fit results or in terms of the coverage in the $\sin^2\theta_{23}-\Delta m^2_{32}$ plane. The fit with event selection shows larger coverage at $99\%$ CL. Note that after event selection the sample size was reduced by $40\%$, which lead to larger statistical uncertainty resulting in worse precision in determining $\sin^2\theta_{23}$ and $\Delta m^2_{32}$ plane. Hence, the rest of the analysis in this paper mainly focus on the fits and the effects of fluctuations without event selection. 

\begin{table}[!htb]
\caption{\label{tab6}Best fit values of the parameters obtained from the fit to five year pseudo-data with (WS) and without (WOS) event selection, for the input values $\sin^2\theta_{23} = 0.5$ and $\Delta m^2_{32} = 2.32\times 10^{-3}~\mathrm{eV^2}$}
\begin{ruledtabular}
\begin{tabular}{lcc}
Parameter & Best-fit value WOS & Best-fit value WS\\
\hline
 $\sin^2\theta_{23}$ & $0.586^{+0.060}_{-0.093}$ & $0.676^{+0.063}_{-0.072}$\\
 $\Delta m^2_{32}$ $(\mathrm{eV^2})$ & $(2.381^{+0.11}_{-0.39})\times10^{-3}$ & $(2.184^{+0.23}_{-0.37})\times10^{-3}$\\
 $f$ & $0.26\pm0.01$ & $0.27\pm0.01$ \\
 $R$ & $5901\pm79$ & $3628\pm62$\\
\end{tabular}
\end{ruledtabular}
\end{table}
\subsection{Effect of fluctuations}
\begin{figure}[!htb]
\subfloat[\label{fig8a}]{%
  \includegraphics[height=0.38\linewidth,width=0.495\linewidth]{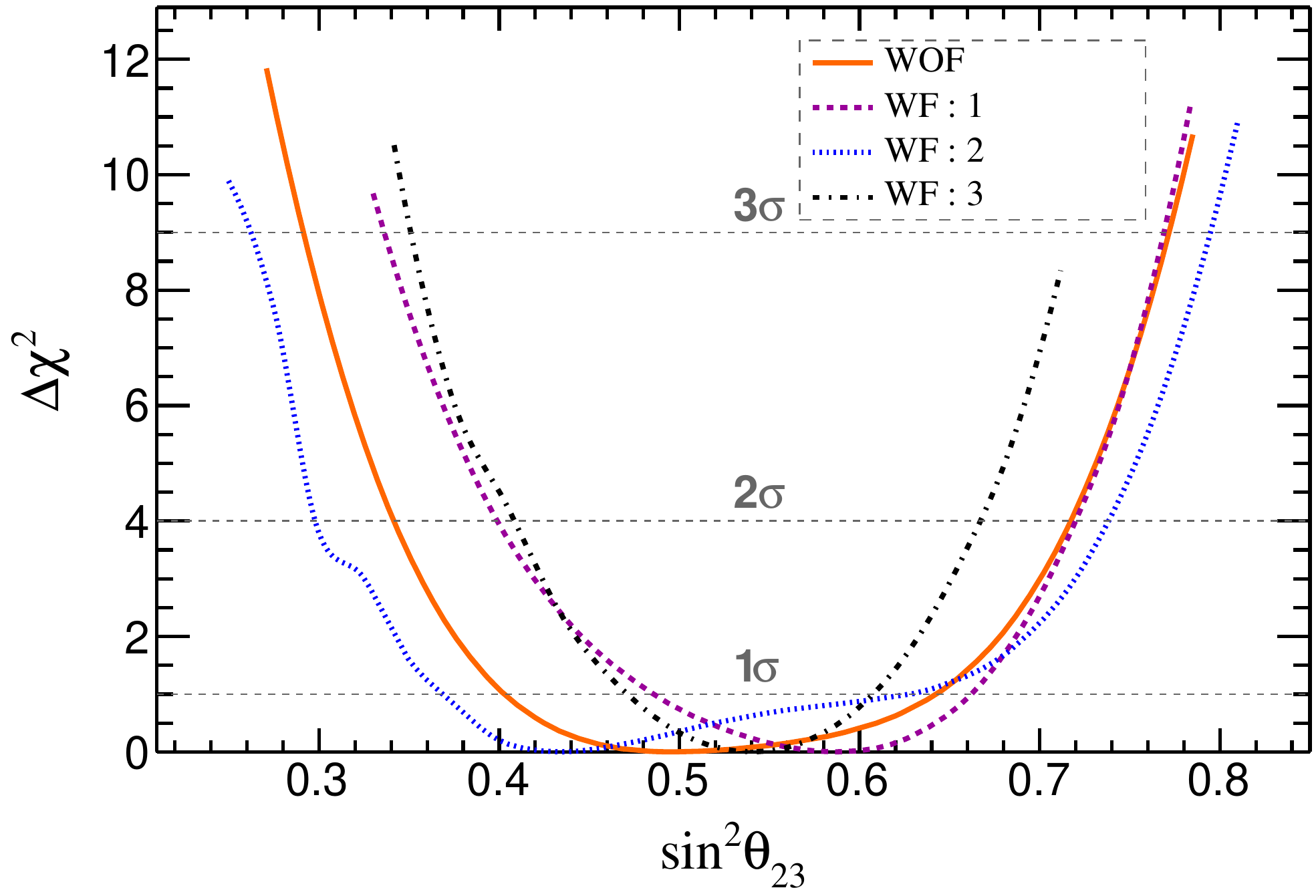}%
}\hfill
\subfloat[\label{fig8b}]{%
  \includegraphics[height=0.38\linewidth,width=0.495\linewidth]{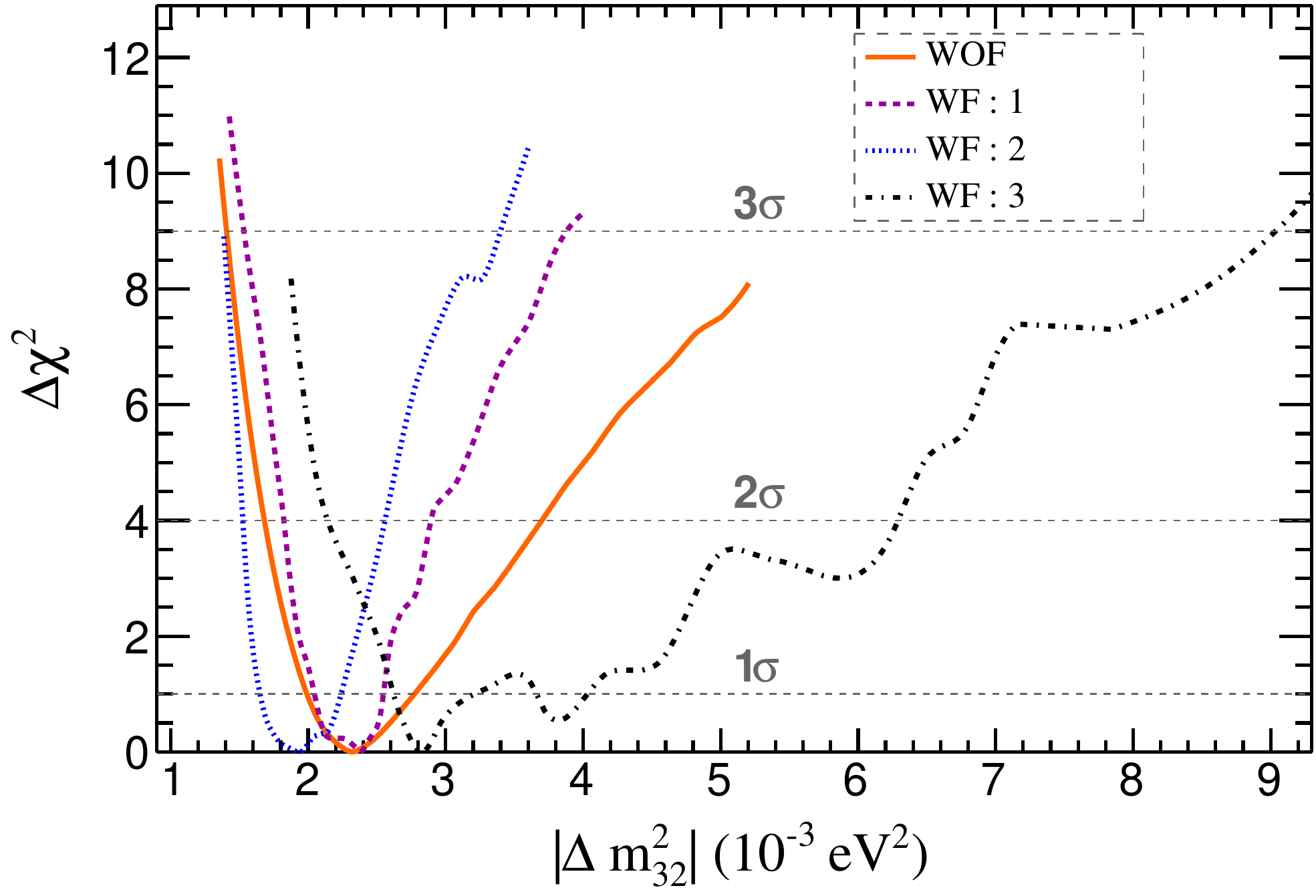}%
}
\caption{\label{fig8}Effect of fluctuations on $\Delta\chi^2$ as a function of \protect\subref{fig8a} $\sin^2\theta_{23}$, for an input value of $\sin^2\theta_{23}(\hbox{true}) = 0.5$ and \protect\subref{fig8b} $\Delta m^{2}_{32}$, for an input value of $\Delta m^{2}_{32}(\hbox{true})= 2.32\times 10^{-3}~\mathrm{eV^2}$ . The solid (orange) curve represents the fit to data without (WOF) fluctuations, and is compared to the fit to two other independent pseudo-data sets with (WF) fluctuations [WF : 1 (dashed, blue) and WF~:~2 (dotted, black)].}
\end{figure}

Earlier analyses \cite{INO_phy} scaled the 1000 year sample to a size corresponding to 5 years to generate the pseudo-data set. The process of scaling nullifies the effect of fluctuations so that the best-fit is always close to the input value. Then the parameter sensitivities are to be understood as the median value when averaged over a large number of randomly generated samples. In order to see this, we generate an unfluctuated 5 year sample by scaling the 1000 year set and a similar analysis is performed as in Section \ref{sec3}. The comparison of $\Delta\chi^2$ with (WF) and without (WOF) fluctuations is shown as a function of $\sin^2\theta_{23}$ and $|\Delta m^2_{32}|$ in Fig. \subref*{fig8a} and \subref*{fig8b} respectively. The ideal fit without fluctuations (WOF) converges near the true (input) value \ie $0.496^{+0.228}_{-0.096}$ in $\sin^2\theta_{23}$ and $(2.32^{+0.43}_{-0.32})\times10^{-3}~\mathrm{eV^2}$ in $\Delta m^2_{32}$. Note that three fits with fluctuations (WF :1, WF : 2 and WF : 3) from three independent fluctuated data sets are used in comparison, and each of them differ in the parameter sensitivities and the best fit values.

\begin{figure}[!htb]
\includegraphics[height=0.5\linewidth,width=0.7\linewidth]{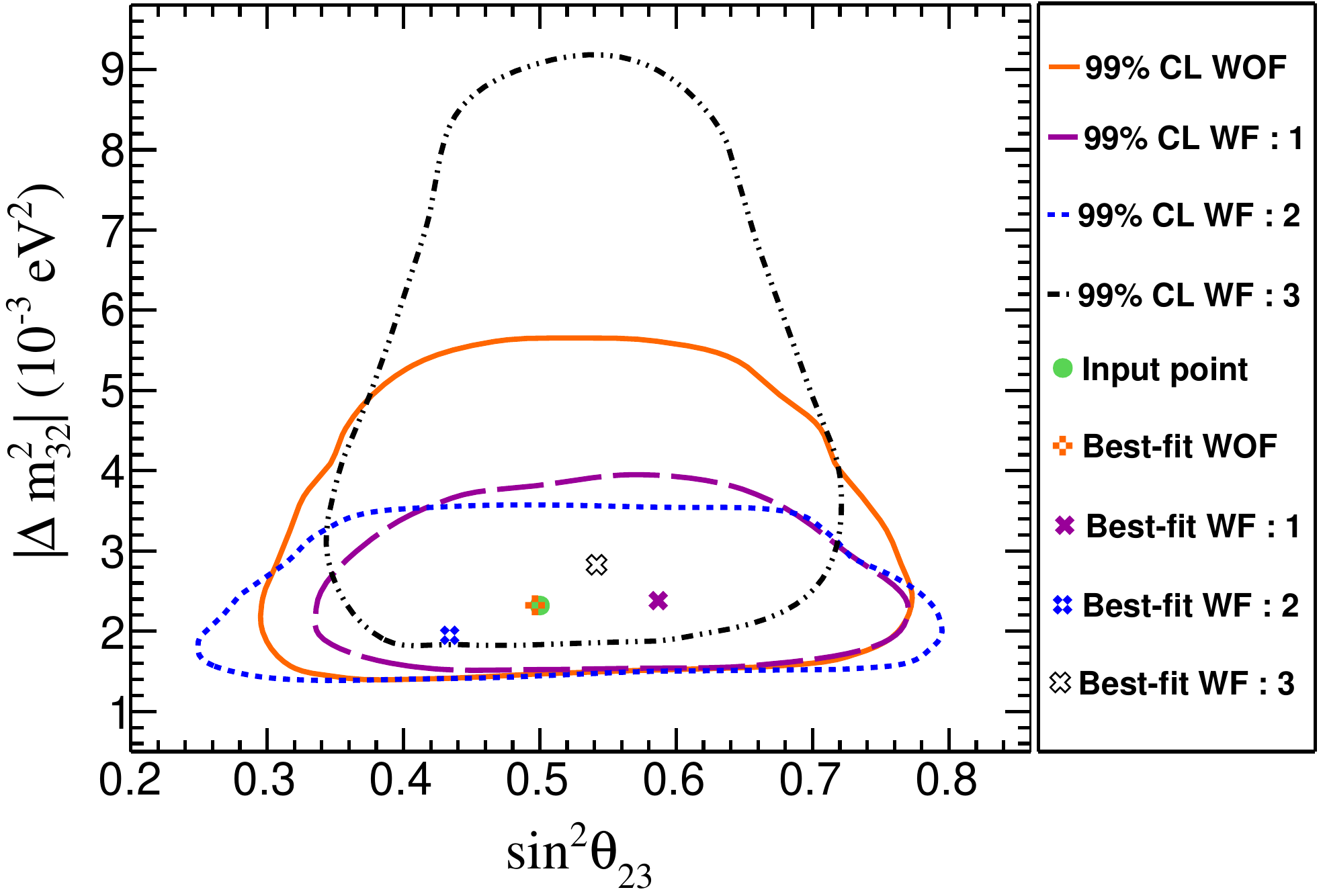}%
\caption{Precision reach obtained from the fit to five year pseudo-data in $\sin^2\theta_{23}-\Delta m^2_{32}$ plane. The solid (orange) line shows the coverage area with $99\%$ CL without (WOF) fluctuations and is compared to the fit to three other independent pseudo-data sets with (WF) fluctuations [WF : 1 (dashed, magenta), WF : 2 (dotted, blue) and WF : 3 (broken, black)]. The input (true) point is given by the green dot.\label{fig9}}
\end{figure}

At maximal mixing in $\sin^2\theta_{23}$, the earth matter effects in atmospheric neutrino oscillation gives better precision in lower octant, which is evident from the smaller uncertainty in the lower octant for the fit without fluctuations. Fluctuations in the data leads to the fluctuations in the octant sensitivities (see Figure \subref*{fig8a}). Also the uncertainties in the parameter determination changes along with the significance of convergence, with each independent fluctuated pseudo-data set. The comparison of precision in $\sin^2\theta_{23}-\Delta m^2_{32}$ plane is shown in Fig. \ref{fig9}, where the fit with fluctuations (WF) shows different coverages in $\sin^2\theta_{23}-\Delta m^2_{32}$ plane for each of the pseudo-data set.

\begin{figure}[!htb]
  \includegraphics[height=0.5\linewidth,width=0.7\linewidth]{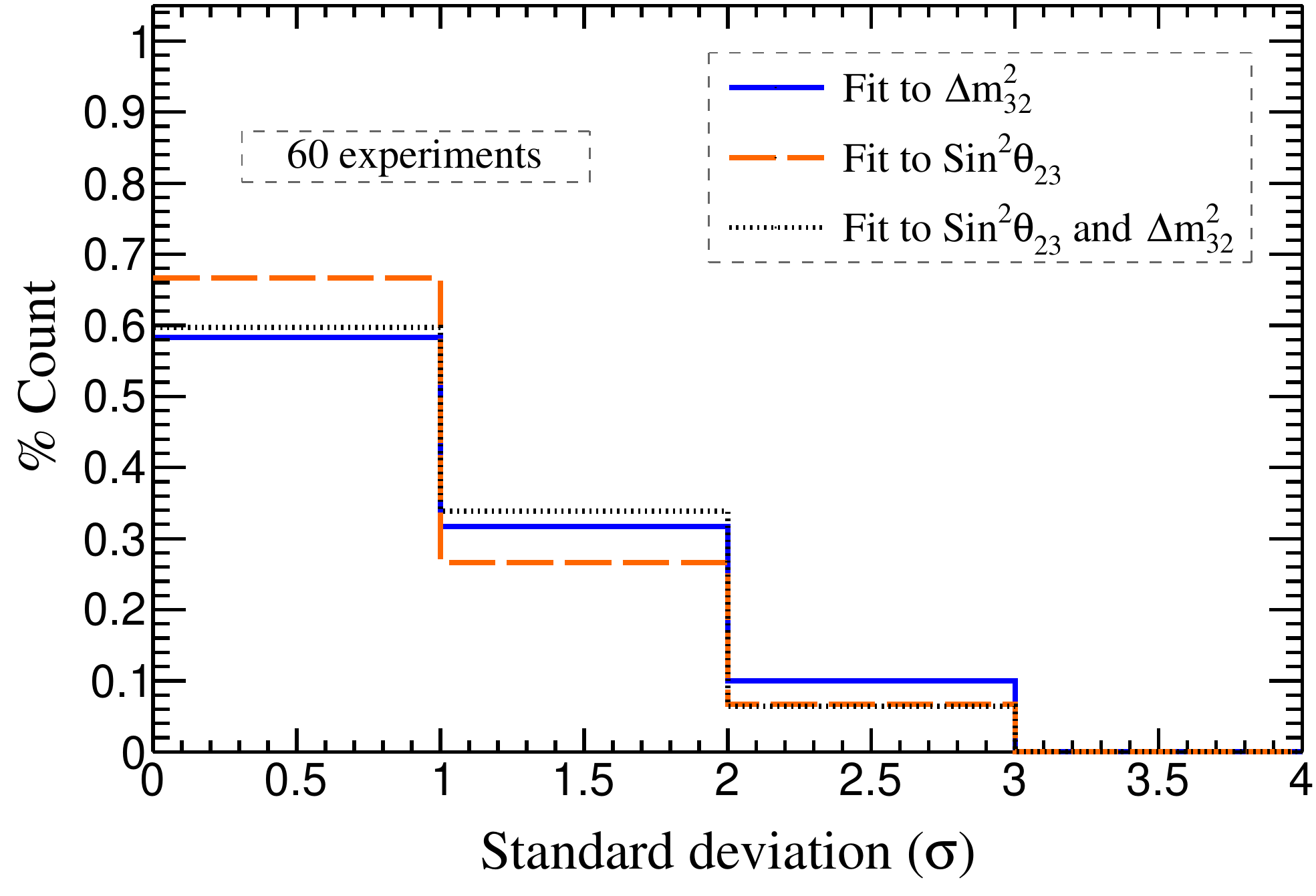}%

\caption{\label{fig10}Significance of convergence obtained from sixty different data sets. The solid (blue) and the broken (orange) line represents the significance for the fit to $\sin^2\theta_{23}$ and $\Delta m^2_{32}$ respectively and the dotted (black) line shows the significance obtained for the simultaneous fit to $\sin^2\theta_{23}$ and $\Delta m^2_{32}$.}
\end{figure}

\begin{figure}[!htb]
  \includegraphics[height=0.5\linewidth,width=0.7\linewidth]{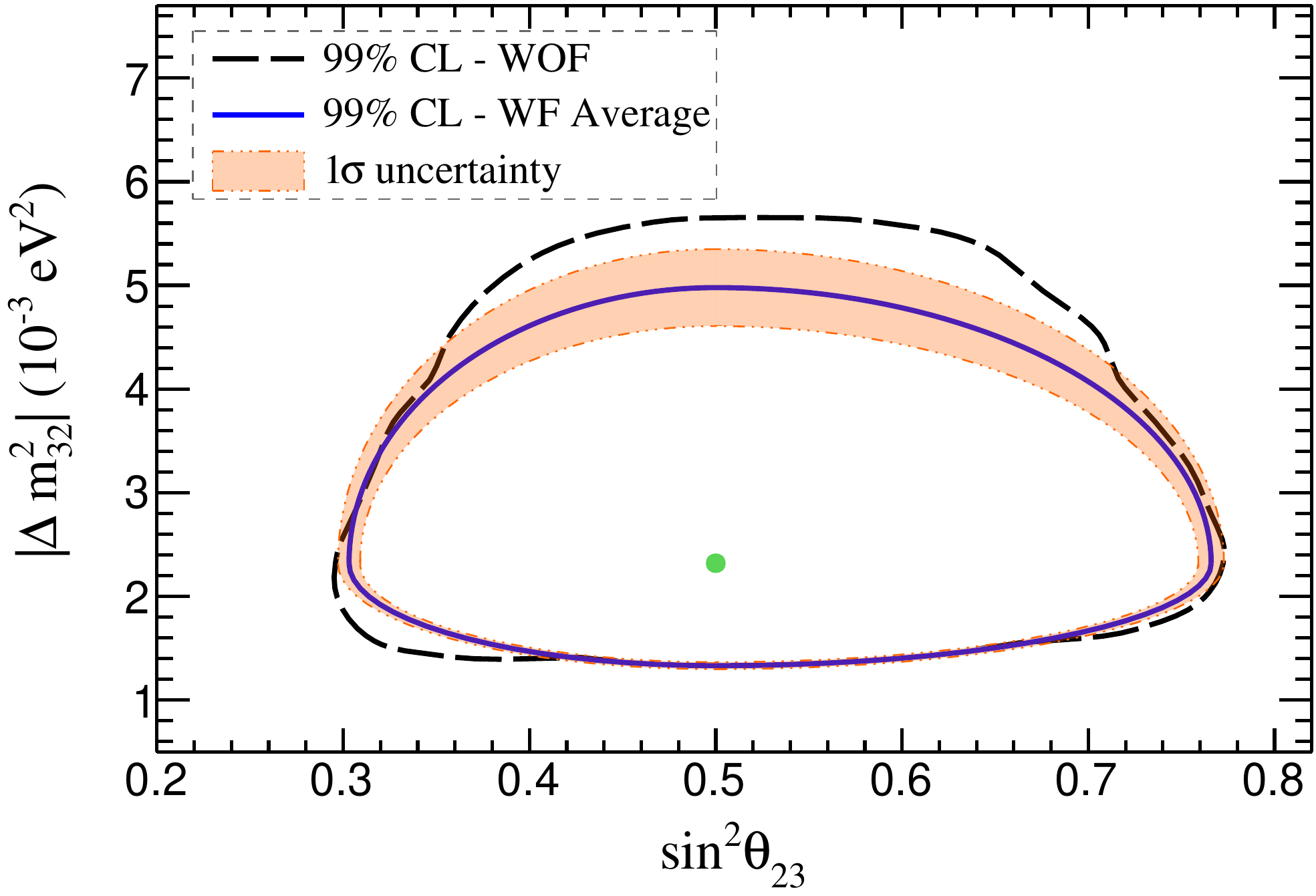}%
\caption{\label{fig11}Comparison of unfluctuated precision reach and the average coverage area calculated from 99\% CL coverages of 50 different sets, in $\sin^2\theta_{23}-\Delta m^2_{32}$ plane. The dashed (black) and solid (blue) line shows the unfluctuated precision reach and average  coverage area respectively with $99\%$ CL. The orange band shows the $1\sigma$ uncertainty in the average coverage and the dot (green) shows the input point.}
\end{figure}

The analysis was repeated for sixty different fluctuated data sets, performing separate (one parameter) and simultaneous (two parameter) fits to determine $\sin^2\theta_{23}$ and $|\Delta m^{2}_{32}|$. Figure \ref{fig10} shows the significance of convergence in terms of standard deviation $\sigma$. As expected, it converges $68\%$ of time within $1\sigma$ of the input value of $\sin^2\theta_{23}=0.5$ and converges within $1\sigma$ of the input value of $\Delta m^{2}_{32}=2.32\times 10^{-3}~\mathrm{eV^2}$ $59\%$ of time for the fit to $\sin^2\theta_{23}$ and $|\Delta m^{2}_{32}|$ respectively. Also $95\%$ of the time the fit converges within $2\sigma$, which evidently shows the Gaussian nature of the fit. The simultaneous fit to $\sin^2\theta_{23}$ and $\Delta m^{2}_{32}$ also shows a similar behaviour in significance. Fig. \ref{fig11} shows the average coverage area with $99\%$ CL in $\sin^2\theta_{23}-\Delta m^2_{32}$ plane, obtained by averaging the coverages from simultaneous fit to fifty different pseudo-data sets. The orange band signifies the $1\sigma$ uncertainty in calculating the average, where the asymmetrical widths from the best fit point of each data set were used. The precision reach for the fit without fluctuations is within $1\sigma$ of the average coverage area calculated.

\begin{figure}[!htb]
\subfloat[\label{fig12a}]{%
  \includegraphics[height=0.38\linewidth,width=0.495\linewidth]{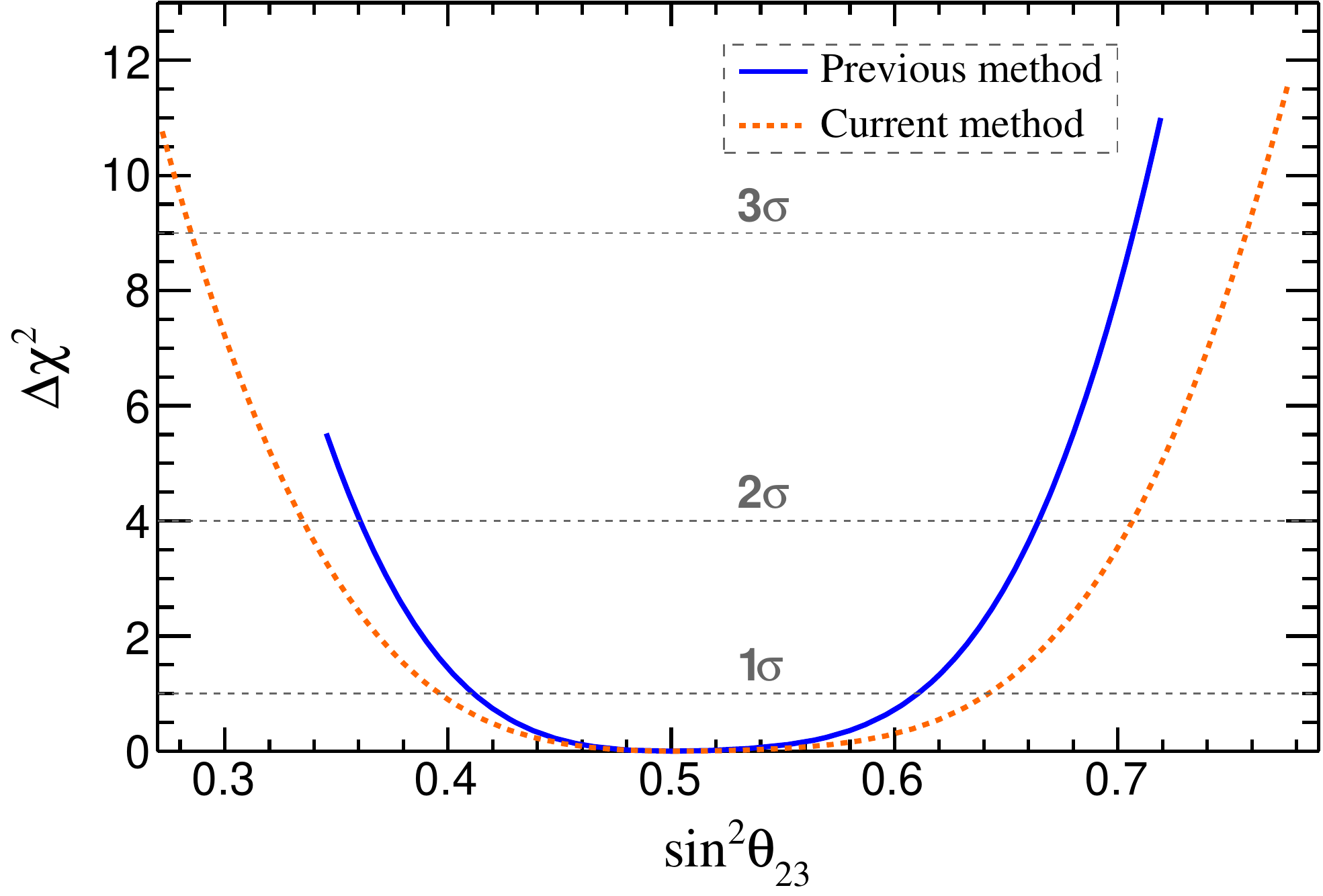}%
}\hfill
\subfloat[\label{fig12b}]{%
  \includegraphics[height=0.38\linewidth,width=0.495\linewidth]{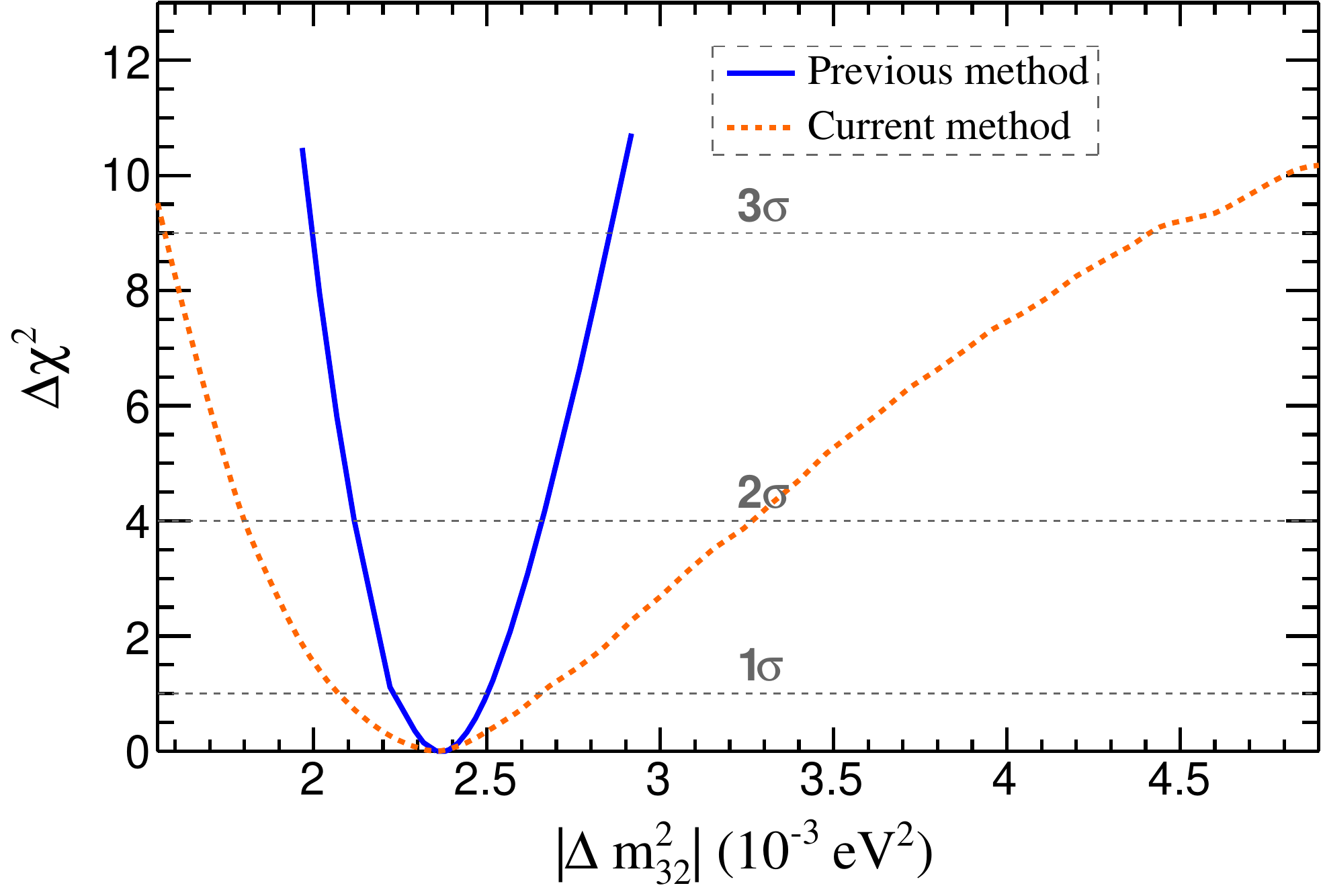}%
}
\caption{\label{fig12}$\Delta\chi^2$ as a function of \protect\subref{fig12a}$\sin^2\theta_{23}$, for an input value of $\sin^2\theta_{23}(\hbox{true}) = 0.5$ and \protect\subref{fig12b} $\Delta m^{2}_{32}$, for an input value of $\Delta m^{2}_{32}(\hbox{true})= 2.32\times 10^{-3}~\mathrm{eV^2}$. The dashed (orange) and solid (blue) line shows the fit obtained from the current and previous methods respectively. Note that the fits are for the same set of NUANCE data using the same binning scheme.}
\end{figure}

Previous studies \cite{INO_phy,Ino_lak} have obtained a better precision in $\sin^2\theta_{23}$ and $|\Delta m^{2}_{32}|$, and have quantified the precision on these parameters as
\begin{equation}
\mathrm{precision}=\frac{P_{\mathrm{max}}-P_{\mathrm{min}}}{P_{\mathrm{max}}+P_{\mathrm{min}}},
\end{equation}
where $P_{\mathrm{max}}$ and $P_{\mathrm{min}}$ are the maximum and minimum values of the concerned  parameter determined at the given C.L. Figure \ref{fig12} compares the $\Delta\chi^2$ obtained from the previous \cite{Ino_lak} and the current analysis methods.  It is to be noted that we have used the same binning scheme and the same set of NUANCE data in both the analysis methods for comparison, but the previous method incorporates the smearing of resolution functions whereas the current method incorporates event-by-event reconstruction. The precision in $\sin^2\theta_{23}$ for the previous method is $19.4\%$ at $1\sigma$, whereas for the current method it deteriorates to $23.8\%$ for five year run of the experiment (see Fig \subref*{fig12a}). The parameter $|\Delta m^2_{32}|$ also shows a similar behavior, where the precision deteriorates from $5.9\%$ to $12.9\%$ at $1\sigma$ for the current method (see Fig \subref*{fig12b}). The drop in precision for the current method is more predominant in $|\Delta m^2_{32}|$ and is clearly seen with a $30\%$ difference in precision at $3\sigma$. 

The noted difference in precision is due to the realistic approach of the event-by-event reconstruction, where the tails of the resolution functions, which were approximated in the previous studies, have been included. In the previous methods, the NUANCE data was folded with the detector efficiencies and were smeared by the resolution functions obtained from GEANT-based studies of single muons with fixed energy and direction \cite{INOmu}. 

\subsection{Mass hierarchy determination}
\begin{figure}[!htb]
\subfloat[\label{fig13a}]{%
  \includegraphics[height=0.38\linewidth,width=0.495\linewidth]{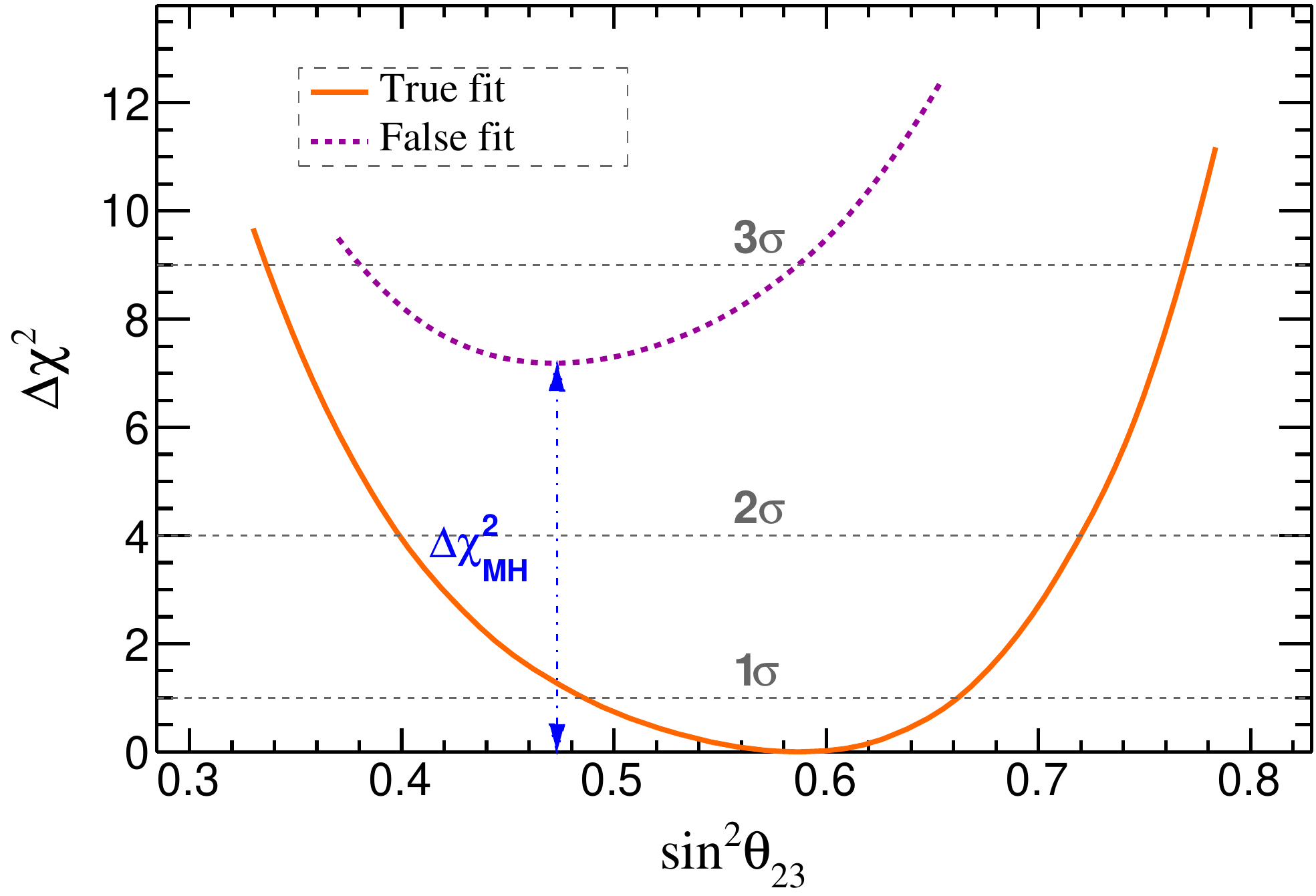}%
}\hfill
\subfloat[\label{fig13b}]{%
  \includegraphics[height=0.38\linewidth,width=0.495\linewidth]{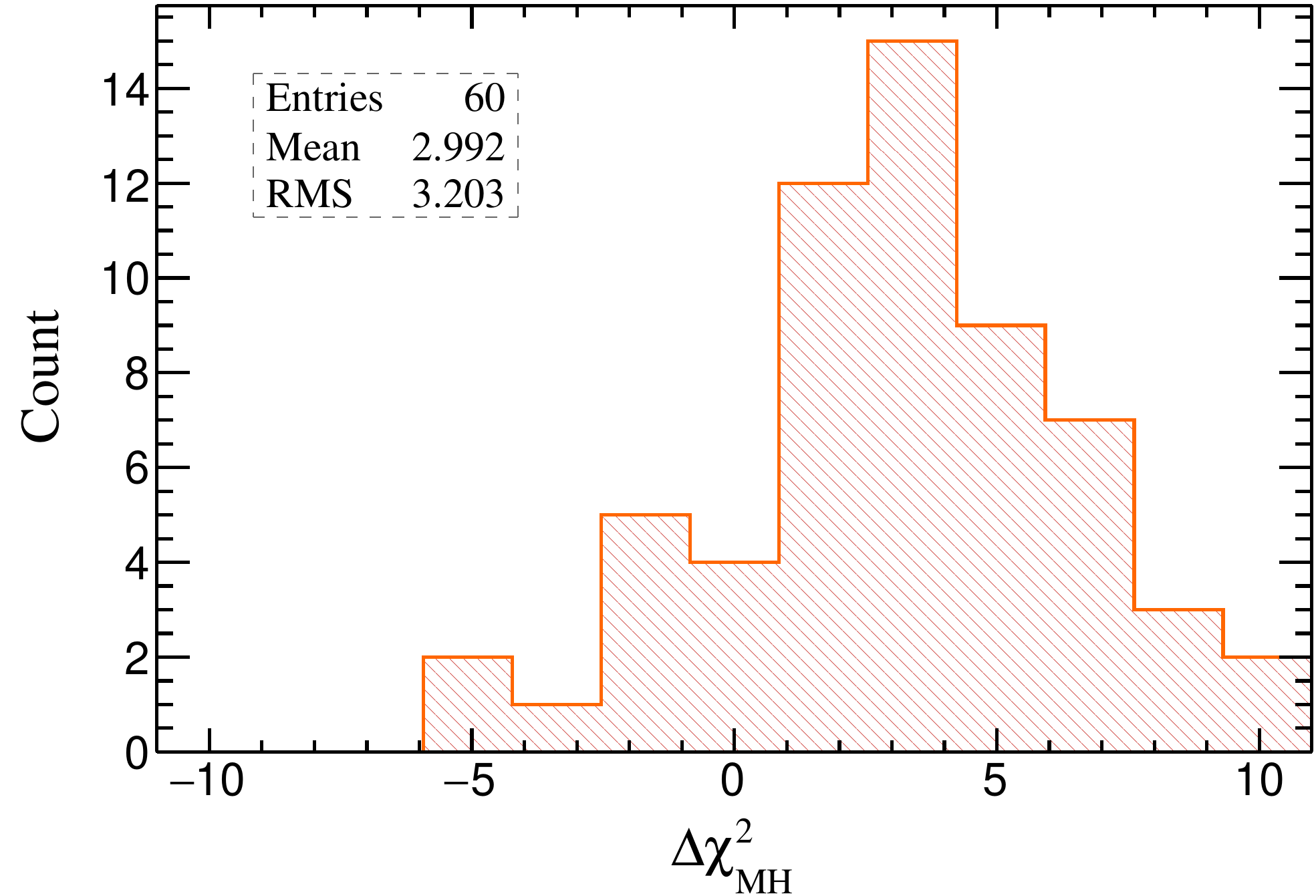}%
}
\caption{\label{fig13}\protect\subref{fig12a}$\Delta\chi^2$ as a function of $\sin^2\theta_{23}$ for true and false fit and \protect\subref{fig12b} Distribution of $\Delta\chi^2_{\mathrm{MH}}$ obtained from fit to sixty fluctuated data sets.}
\end{figure}

The five year pseudo-data set is oscillated via accept or reject method assuming NH (IH), which is then fit with true NH (IH) and false IH (NH) PDFs. The parameters in the fit are marginalized as given in Table \ref{tab5}, and the $\Delta\chi^2$ resolution to differentiate the correct hierarchy from the wrong hierarchy is defined as:
\vspace*{-0.3cm}
\begin{equation}
\Delta\chi^2_{\mathrm{MH}}=\chi^2_{\mathrm{false}}-\chi^2_{\mathrm{true}}~,
\end{equation}
where $\chi^2_{\mathrm{true}}$ and $\chi^2_{\mathrm{false}}$ are the minimum values of  $\chi^2$ from the true and false fits respectively. Fig. \subref*{fig13a} shows the true and false hierarchical fit to $\sin^2\theta_{23}$ for a particular pseudo-data set with fluctuations, wherein the resolution of $\Delta\chi^2_{\mathrm{MH}}= 7.2$ rules out the wrong hierarchy with a significance greater than $2\sigma$ for this set.

\paragraph*{}
The procedure was repeated for sixty independent five-year fluctuated data sets to see the effect of fluctuations on the mass hierarchy significance. Figure \subref*{fig13b} shows the distribution of $\Delta\chi^2_{\mathrm{MH}}$ obtained from the fit to sixty different sets. The mean resolution of  $\Delta\chi^2_{\mathrm{MH}}=2.9$ rules out the wrong hierarchy with a significance of $\approx 1.7 \sigma$ for a five year run of 50 kton ICAL detector. The large uncertainty in $\Delta\chi^2_{\mathrm{MH}}$ is due to the fluctuations in the data and the negative values signifies the identification of the wrong mass hierarchy. Note that an earlier analysis that excluded the effect of fluctuations \cite{Ino_lak} gave a value $\Delta\chi^2_{\mathrm{MH}}=2.7$ and our mean value is compatible with this, as expected, given minor differences in the analysis procedures.

\section{Discussion and Summary}\label{sec5}
One of the main aims of the proposed ICAL at INO is to measure the atmospheric neutrino oscillation parameters $\sin^2\theta_{23}$ and $|\Delta m^{2}_{32}|$, and also to measure the mass hierarchy (MH) of neutrinos. The moderately large value of $\theta_{13}$ and the ability of the magnetised ICAL to distinguish a neutrino event form an antineutrino event, allows the observation of  earth matter effects separately in $\nu$ and $\bar\nu$ and helps identify the MH of neutrinos. In this analysis we focus on the precision measurements and the mass hierarchy resolutions that ICAL could attain within a period of five years.

\paragraph*{}
Incorporating a realistic analysis procedure, it is for the first time we have applied event-by-event reconstruction and have considered the tails of angular and energy resolution which were approximated by single Gaussians and Vavilov functions in previous such studies \cite{INO_phy}. We show that incorporating non-Gaussian resolutions are likely to effect the parameter sensitivities. Also for the first time we study the effect of low event statistics on the precision and MH measurements, by introducing fluctuations in the data. It is also for the first time within the framework of low event statistics, we show the effect of event selection criterion on the parameter sensitivities, and show that we can include all reconstructed muons to get better sensitivity of parameters. Hence within the framework of low event statistics, we show that the fit without any selection criterion (WOS), where we include all the reconstructed events, is the baseline to obtain a better constraints on the parameters.

\paragraph*{}
We start by using five year fluctuated pseudo-data set for our analysis and apply oscillations via the accept or reject method. The oscillated data is binned in $E_{\mu}$ and $\cos\theta_z$ for the $\chi^2$ analysis, where we have used the energy and direction information of the muon from the CC $\nu_\mu$ events only. The constraints on $\sin^2\theta_{23}$ and $\Delta m^{2}_{32}$ are compared with and without event selection criterion. Statistically we loose $40\%$ of events after selection, hence we find large uncertainty in parameter determination after applying event selection. We use an ensemble of independent fluctuated data sets to study the effect of low event statistics on the precision measurements of the oscillation parameters. The constrains on $\sin^2\theta_{23}$ and $\Delta m^{2}_{32}$ are compared with and without fluctuations, and we find a reasonable agreement between the unfluctuated and the average fluctuated precision reach obtained in $\sin^2\theta_{23}-\Delta m^{2}_{32}$ plane.
\paragraph*{}
As far as the mass hierarchy of the neutrinos is concerned, we find a mean resolution of $\Delta\chi^2_{MH}=2.9$ from an ensemble of sixty experiments. This rules out the wrong hierarchy with a significance of $\approx1.7\sigma$, consistent with earlier analysis obtained without considering fluctuations. We also find a significant deviation in the mean value of $\Delta\chi^2_{MH}$, and roughly a $15\%$ probability of obtaining the wrong hierarchy due to the fluctuations in the data.
\paragraph*{}
This paper presents an analysis procedure which can be used on the real ICAL data where the fluctuations are inbuilt as the PDFs are uncorrelated. However, in this analysis we have only used muon information from CC $\nu_\mu$ events and we have ignored the small contribution from $\nu_e$ to $\nu_{\mu}$ oscillated events, that is known to slightly dilute the sensitivities \cite{Indu_prem}. The ICAL can also measure the hadron energy via proper calibration of hits, and including the hadron energy information in CC events is expected to improve the sensitivity of the detector towards the oscillation parameters and also improve the MH significance \cite{INOhad}. Note that there are CC $\nu_e$ as well as neutral current (NC) events in the detector. Separation of $\nu_{\mu}$ CC events from the others is quite robust for $E_{\mu}\gtrsim 1 \mathrm{GeV}$ and has been discussed elsewhere \cite{INO_phy}. Separation of low energy CC $\nu_{\mu}$ events from CC $\nu_e$ and NC events is an ongoing effort of the INO-ICAL collaboration. A combined analysis including all the CC events along with the hadron information will give us the maximum sensitivity the ICAL can attain, and is likely to improve the results presented in this paper.

\begin{acknowledgments}
We thank G. Majumder for providing us the ICAL detector simulation package. We are also thankful for the excellent computing facilities and support from the P.G. Senapathy centre for computing resource at IITM, Chennai and the computing facility at TIFR, Mumbai, which made the intensive computation required for this analysis possible. 
\end{acknowledgments}

\bibliographystyle{apsrev4-1}
\bibliography{document}

\end{document}